\documentclass[11pt,a4paper]{article}
\usepackage[utf8]{inputenc}

\usepackage{authblk}
\usepackage{amsmath}
\usepackage{amssymb}

\usepackage{color}

\usepackage{bm}

\newcommand{\pn}{\mathrm{P}}
\newcommand{\intn}{\mathrm{I}}
\newcommand{\leak}{\mathrm{lk}}
\newcommand{\dnd}{\mathrm{D}}
\newcommand{\apic}{\mathrm{A}}
\newcommand{\bas}{\mathrm{B}}
\newcommand{\som}{\mathrm{som}}
\newcommand{\exc}{\mathrm{exc}}
\newcommand{\inh}{\mathrm{inh}}
\newcommand{\rest}{\mathrm{rest}}
\newcommand{\topdown}{\mathrm{TD}}

\DeclareMathOperator{\Tr}{Tr}
\DeclareMathOperator{\Expect}{E}

\usepackage{jneurosci}

\usepackage{setspace}
\usepackage[a4paper, top=1in, bottom=1in, left=1in, right=1in]{geometry}

\usepackage[scaled]{helvet}
\usepackage{sfmath}

\newcommand{\beginsupplement}{%
        \setcounter{table}{0}
        \renewcommand{\thetable}{S\arabic{table}}%
        \setcounter{equation}{0}
        \renewcommand\theequation{S\arabic{equation}}
        \setcounter{figure}{0}
        \renewcommand{\thefigure}{S\arabic{figure}}%
     }

\usepackage{graphicx}

\renewenvironment{abstract}
 {\begin{center}
  \bfseries \abstractname\vspace{-.5em}\vspace{0pt}
  \end{center}
  \list{}{%
    \setlength{\leftmargin}{2mm}%
    \setlength{\rightmargin}{\leftmargin}%
  }%
  \item\relax}
 {\endlist}

\title{\vspace{-5mm}Dendritic error backpropagation\\ in deep cortical microcircuits}
\author[1*]{João Sacramento}
\author[1]{Rui Ponte Costa}
\author[2]{Yoshua Bengio}
\author[1*]{Walter Senn}
\affil[1]{Department of Physiology\protect\\University of Bern, Switzerland\vspace{2mm}}
\affil[2]{Montreal Institute for Learning Algorithms\protect\\Université de Montréal, Quebec, Canada}

\begin{document}

\date{}

\maketitle
\begin{abstract}
\normalsize
Animal behaviour depends on learning to associate sensory stimuli with the desired motor command.
Understanding how the brain orchestrates the necessary synaptic modifications across different brain areas has remained a longstanding puzzle.
Here, we introduce a multi-area neuronal network model in which synaptic plasticity continuously adapts the network towards a global desired output. In this model synaptic learning is driven by a local dendritic prediction error that arises from a failure to predict the top-down input given the bottom-up activities. Such errors occur at apical dendrites of pyramidal neurons where both long-range excitatory feedback and local inhibitory predictions are integrated. When local inhibition fails to match excitatory feedback an error occurs which triggers plasticity at bottom-up synapses at basal dendrites of the same pyramidal neurons. We demonstrate the learning capabilities of the model in a number of tasks and show that it approximates the classical error backpropagation algorithm. Finally, complementing this cortical circuit with a disinhibitory mechanism enables attention-like stimulus denoising and generation. Our framework makes several experimental predictions on the function of dendritic integration and cortical microcircuits, is consistent with recent observations of cross-area learning, and suggests a biological implementation of deep learning.
\let\thefootnote\relax\footnotetext{* Corresponding authors: \{sacramento\},\{senn\}@pyl.unibe.ch}
\end{abstract}

\doublespacing

\section*{Introduction}

While walking on the street we are constantly bombarded with complex sensory stimuli.
Learning to navigate such complex environments is of fundamental importance for survival. In the brain, these forms of learning are believed to rely on the orchestrated wiring of synaptic communication between different cortical areas, such as visual and motor cortices \cite{Petreanu:2012dy,Manita2015,Makino:2015jr,Poort:2015fwa,Zmarz:2016jo,Attinger:2017bo}. However, how to correctly modify synapses to achieve an appropriate interaction between brain areas has remained an open question. This fundamental issue in learning and development is often referred to as the credit assignment problem \cite{Rumelhart1986,sutton1998reinforcement,Roelfsema2005,Friedrich:2011jm,Bengio2014}. The brain, and artificial neural networks alike, have to determine how to best modify a given synapse across multiple processing stages to ultimately improve global behavioural output.

Machine learning has recently undergone remarkable progress through the use of deep neural networks, leading to human-level performance in a growing number of challenging problems \cite{LeCun2015}. Key to an overwhelming majority of these achievements has been the backpropagation of errors algorithm (backprop; \citenoparens{Rumelhart1986}), which has been long dismissed in neuroscience on the grounds of biologically implausibility \cite{Grossberg1987,Crick1989}. Nonetheless, accumulating evidence highlights the difficulties of simpler learning models and architectures in accurately reproducing cortical activity patterns when compared to deep neural networks, notably trained only on sensory data \cite{Yamins2014,Ravazi2014,Yamins2016}. Although recent developments have started to bridge the gap between neuroscience and artificial intelligence \cite{Marblestone:2016bm,Lillicrap2016,costa2017cortical,Guerguiev2017}, whether the brain implements a backprop-like algorithm remains unclear.

Here we propose that the errors at the heart of backprop are encoded on the distal dendrites of cross-area projecting pyramidal neurons. In our model, these errors arise from a failure to exactly match via lateral (e.g.~somatostatin-expressing, SST) interneurons the top-down feedback from downstream cortical areas. Synaptic learning is driven by these error-like signals that flow through the dendrites and trigger plasticity on bottom-up connections. Therefore, in contrast to previous approaches \cite{Marblestone:2016bm}, in our framework a given neuron is used simultaneously for activity propagation (at the somatic level), error encoding (at distal dendrites) and error propagation to the soma. Importantly, under certain simplifying assumptions, we were able to formally show that learning in the model approximates backprop.

We first illustrate the different components of the model and afterwards demonstrate its performance by training a multi-area network on associative nonlinear regression and recognition tasks (handwritten digit image recognition, a standard benchmark). Then, we further extend the framework to consider learning of the top-down synaptic pathway. When coupled with a disinhibitory mechanism this allows the network to generate prototypes of learnt images as well as perform input denoising. We interpret this disinhibitory mechanism as being implemented through another inhibitory cell-type (e.g.~vasoactive intestinal peptide-expressing, VIP interneurons). Finally, we make several experimentally testable predictions in terms of the role of dendrites and different interneuron types being involved while an animal learns to associate signals originating from different brain areas.

\section*{Results}

The cortex exhibits remarkably intricate, but stereotypical circuits. Below we describe a plastic cortical circuit model that considers two features observed in neocortex: dendritic compartments and different cell types. Cross-area synapses onto the dendritic compartments learn to reduce the prediction error between the somatic potential and their own dendritic branch potential. Additionally, lateral synaptic input from local interneurons learns to cancel top-down feedback from downstream brain areas. When a new top-down input arrives at distal dendrites that cannot be matched by lateral inhibition it signals a neuron-specific error (encoded on the dendritic potential) that triggers synaptic learning at a given pyramidal cell. As learning progresses, the interneurons gradually learn to cancel once again the new input until eventually learning stops. We show that this cortical circuit implements error backpropagation, and demonstrate its performance in various tasks.
\subsection*{The dendritic cortical circuit learns to predict self-generated top-down input}

\begin{figure}[htb!]
  \centering
  \includegraphics[width=1\linewidth]{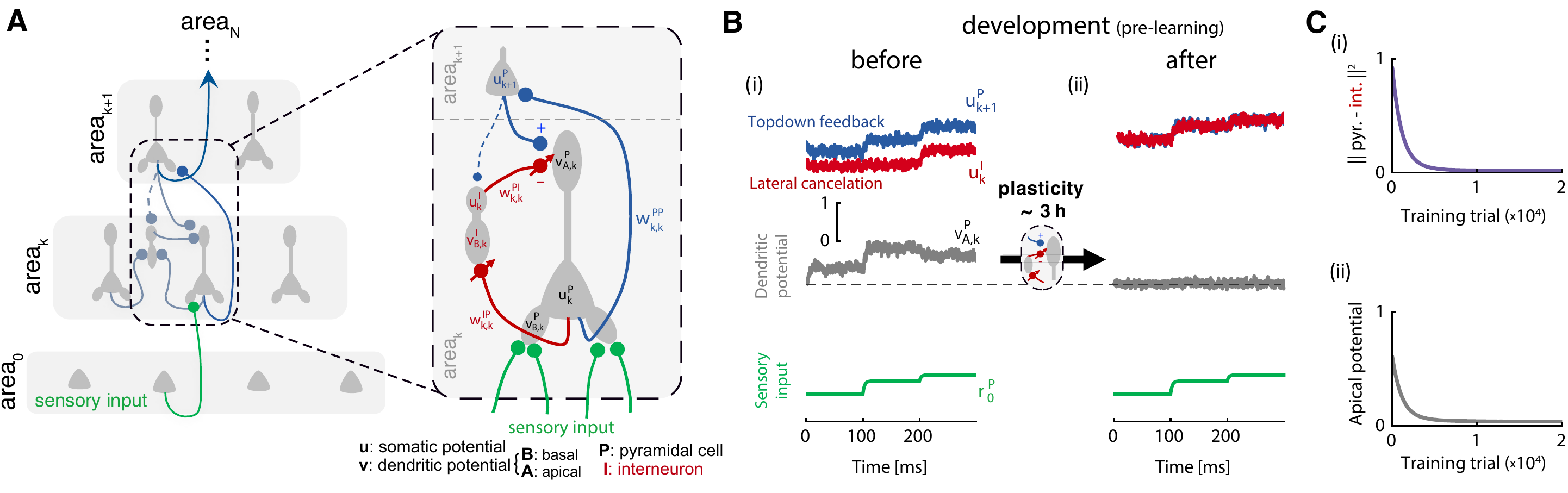}
  \caption{\textbf{Dendritic cortical circuit learns to predict self-generated top-down input.}
  (\textbf{A}) Illustration of multi-area network architecture. The network consists of an input area $0$ (e.g., thalamic input), one or more intermediate (hidden) areas (represented by area $k$ and area $k\!+\!1$, which can be mapped onto primary and higher sensory areas) and an output area $N$ (e.g. motor cortex) (left). Each hidden area consists of a microcircuit with pyramidal cells and lateral inhibitory interneurons (e.g.~somatostatin-positive, SST, cells) (right). Pyramidal cells consist of three compartments: a basal compartment (with voltage $\mathbf{v}_{\bas,k}^\pn$) that receives bottom-up input; an apical compartment ($\mathbf{v}_{\apic,k}^\pn$), where top-down input converges to; and a somatic compartment ($\mathbf{u}_k^\pn$), that integrates the basal and apical voltage. Interneurons receive input from lateral pyramidal cells onto their own basal dendrites ($\mathbf{v}_{\bas,k}^\intn$), integrate this input on their soma ($\mathbf{u}_k^\intn$) and project back to the apical compartments ($\mathbf{v}_{\apic,k}^\pn$) of same-area pyramidal cells.
  (\textbf{B}) In a pre-learning developmental stage, the network learns to predict and cancel top-down feedback given randomly generated inputs. Only pyramidal-to-interneuron synapses ($\mathbf{W}^{\intn\pn}_{k,k}$) and interneuron-to-pyramidal synapses ($\mathbf{W}^{\pn\intn}_{k,k}$) are changed at that stage according to predictive synaptic plasticity rules (defined in Eqs \ref{eq:dW-IP} and \ref{eq:dW-PI} of the Methods). Example voltage traces for a randomly chosen downstream neuron ($\mathbf{u}^\pn_{k+1}$) and a corresponding interneuron ($\mathbf{u}^\intn_k$), a pyramidal cell apical compartment ($\mathbf{v}_{\apic,k}^\pn$) and an input neuron ($\mathbf{u}_0^\pn$), before (i) and after (ii) development, for three consecutively presented input patterns. Once learning of the lateral synapses from and onto interneurons has converged, self-generated top-down signals are predicted by the network --- it is in a \emph{self-predicting state}. Here we use a concrete network with one hidden area and 30-20-10 pyramidal neurons (input-hidden-output), but the particular network dimensions do not impact the ability of the network to produce these results. Note that no desired targets are presented to the output area (cf.~Fig.~\ref{fig:learning_newinput}); the network is solely driven by random inputs.
  (\textbf{C}) Lateral inhibition cancels top-down input. (i) Interneurons learn to match next-area pyramidal neuron activity as their input weights $\mathbf{W}^{\intn\pn}_{k,k}$ adapt (see main text and Methods for details). (ii) Concurrently, learning of interneuron-to-pyramidal synapses ($\mathbf{W}^{\pn\intn}_{k,k}$) silences the apical compartment of pyramidal neurons, but pyramidal neurons remain active (cf.~B). This is a general effect, as the lateral microcircuit learns to predict and cancel the expected top-down input for every random pattern (see SI).}
  \label{fig:learn-self-predicting-state}
\end{figure}

We first study a generic network model with $N$ cortical brain areas (a multi-layer network, in machine learning parlance), comprising an input area (representing, for instance, thalamic input to sensory areas, denoted as area $0$), one or more `hidden' areas (representing secondary sensory and consecutive higher brain areas, denoted as area $k$ and area $k\!+\!1$, respectively) and an output brain area (e.g. motor cortex, denoted as area $N$), see schematic in Fig.~\ref{fig:learn-self-predicting-state}A. Unlike conventional artificial neural networks, hidden neurons feature both bottom-up ($\mathbf{W}_{k+1,k}^{\pn\pn}$) and top-down ($\mathbf{W}_{k,k+1}^{\pn\pn}$) connections, thus defining a recurrent network structure. Top-down synapses continuously feed back the next brain area predictions to a given bottom-up input. Our model uses of this feedback to determine corrective error signals and ultimately guide synaptic plasticity across multiple areas.

Building upon previous work \cite{Urbanczik2014}, we adopt a simplified  multicompartment neuron and describe pyramidal neurons as three-compartment units (schematically depicted in Fig.~\ref{fig:learn-self-predicting-state}A; see also Methods). These compartments represent the somatic, basal and apical integration zones that characteristically define neocortical pyramidal cells \cite{Spruston2008,Larkum2013}. The dendritic structure of the model is exploited by having bottom-up and top-down synapses converging onto separate dendritic compartments (basal and distal dendrites, respectively), consistent with experimental observations \cite{Spruston2008} and reflecting the preferred connectivity pattern of cortico-cortical projections \cite{Larkum2013}.

Consistent with the neurophysiology of SST interneurons \cite{UrbanCiecko:2016io}, we also introduce a second population of cells within each hidden area with both lateral and cross-area connectivity, whose role is to cancel the top-down input. Modelled as two-compartment units (depicted in red, Fig.~\ref{fig:learn-self-predicting-state}A; see also Methods), such interneurons are predominantly driven by pyramidal cells within the same area through weights $\mathbf{W}_{i,i}^{\intn\pn}$, and they project back to the apical dendrites of the same-area pyramidal cells through weights $\mathbf{W}_{k,k}^{\pn\intn}$ (see Fig.~\ref{fig:learn-self-predicting-state}A). Additionally, cross-area feedback onto SST cells originating at the next higher brain area $k\!+\!1$ provide a weak nudging signal for these interneurons, modelled after \citetext{Urbanczik2014} as a conductance-based somatic input current. For computational simplicity, we modelled this weak top-down nudging on a one-to-one basis (that can also be relaxed): each interneuron is nudged towards the potential of a corresponding upper-area pyramidal cell. Recent monosynaptic input mapping experiments show that somatostatin-positive cells (SST, of which Martinotti cells are the main type) in fact receive also top-down projections \cite{Leinweber:2017jq}, that according to our proposal encode the weak 'teaching' signals from higher to lower brain areas.

As detailed below, this microcircuit is key to encode and backpropagate errors across the network. We first show how synaptic plasticity of lateral interneuron connections establishes a network regime, which we term \emph{self-predicting}, whereby lateral input cancels the self-generated top-down feedback, effectively silencing apical dendrites. For this reason, SST cells are functionally inhibitory and are henceforth referred to as interneurons. Crucially, when the circuit is in this so-called self-predicting state, presenting a novel external signal at the output area gives rise to top-down activity that cannot be explained away by the interneuron circuit. Below we show that these apical mismatches between top-down and lateral input constitute the backpropagated, neuron-specific errors that drive plasticity on the forward weights to the hidden pyramidal neurons.

Learning to predict the feedback signals involves adapting both weights from and to the lateral interneuron circuit. Consider a network that is driven by a succession of sensory input patterns (Fig.~\ref{fig:learn-self-predicting-state}B, bottom row). The exact distribution of inputs is unimportant as long as they span the whole input space (see SI). Learning to cancel the feedback input is divided between both the weights from pyramidal cells to interneurons, $\mathbf{W}_{k,k}^{\intn\pn}$, and from interneurons to pyramidal cells, $\mathbf{W}_{k,k}^{\pn\intn}$.

First, due to the somatic teaching feedback, learning of the $\mathbf{W}_{k,k}^{\intn\pn}$ weights leads interneurons to better reproduce the activity of the respective higher brain area $k\!+\!1$ (Fig.~\ref{fig:learn-self-predicting-state}B (i)). A failure to reproduce area $k\!+\!1$ activity generates an internal prediction error at the dendrites of the interneurons, which triggers synaptic plasticity (as defined by Eq.~\ref{eq:dW-IP} in the Methods) that corrects for the wrong dendritic prediction and eventually leads to a faithful tracing of the upper area activity by the lower area interneurons  (Fig.~\ref{fig:learn-self-predicting-state}B (ii)). The mathematical analysis (see SI, Eq.~\ref{eq:expected-dW-IP}) shows that the plasticity rule \eqref{eq:dW-IP} makes the inhibitory population implement the same function of the area-$k$ pyramidal cell activity as done by the area--($k\!+\!1$) pyramidal neurons. Thus, the interneurons will learn to mimic the area--($k\!+\!1$) pyramidal neurons (Fig.~\ref{fig:learn-self-predicting-state}Ci).

Second, as the interneurons mirror upper area activity, inter-to-pyramidal neuron synapses within the same area ($\mathbf{W}_{k,k}^{\pn\intn}$, Eq.~\ref{eq:dW-PI}) successfully learn to cancel the top-down input to the apical dendrite (Fig.~\ref{fig:learn-self-predicting-state}Cii), independently of the actual input stimulus that drives the network. By doing so, the inter-to-pyramidal neuron weights $\mathbf{W}_{k,k}^{\pn\intn}$ learn to mirror the top-down weights onto the lower area pyramidal neurons. The learning of the weights onto and from the interneurons works in parallel: as the interneurons begin to predict the activity of pyramidal cells in area $k\!+\!1$, it becomes possible for the plasticity at interneuron-to-pyramidal synapses (Eq.~\ref{eq:dW-PI}) to find a synaptic weight configuration which precisely cancels the top-down feedback (see also SI, Eq.~\ref{eq:expected-dW-PI}). At this stage, every pattern of activity generated by the hidden areas of the network is explained by the lateral circuitry, Fig.~\ref{fig:learn-self-predicting-state}C~(ii). Importantly, once learning of the lateral interneurons has converged, the apical input cancellation occurs irrespective of the actual bottom-up sensory input. Therefore, interneuron synaptic plasticity leads the network to a \emph{self-predicting state}. We propose that the emergence of this state could occur during development, consistent with experimental findings \cite{Dorrn:2010hu,Froemke:2015kx}. Starting from a cross-area self-predicting configuration helps learning of specific tasks (but is not essential, see below and Methods).

\subsection*{Deviations from self-predictions encode backpropagating errors}

\begin{figure}[htb!]
  \centering
  \includegraphics[width=1\linewidth]{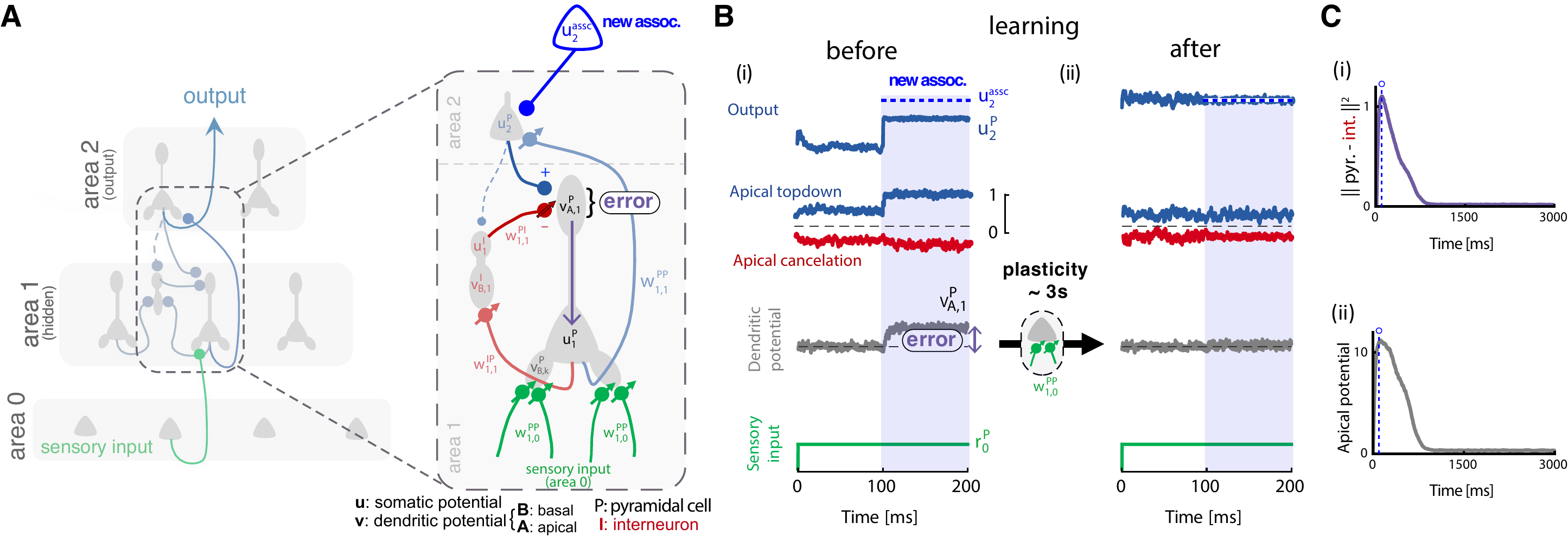}
  \caption{\textbf{Deviations from self-predictions encode backpropagating errors that are used for learning in bottom-up synapses.}
  (\textbf{A}) When a novel associative (or `teaching') signal is presented to the output area ($\mathbf{u}_2^{\mathrm{assoc}}$, blue at the top), a prediction error in the apical compartments of pyramidal neurons in the upstream area (area 1, `error') is generated. This error appears as an apical voltage deflection that propagates down to the soma (purple arrow) where it modulates the somatic firing rate. Bottom-up synapses at the basal dendrites learn to predict the somatic firing rate (bottom, green). Only the elements directly involved in encoding the error and modifying the bottom-up synapses are highlighted in this microcircuit.
  (\textbf{B}) Activity traces in the microcircuit before and after a new associative signal is learned. (i) Before learning, when a new associative signal is presented at the output area  ($\mathbf{u}_2^\mathrm{assoc}$, blue dashed), the somatic voltage of output neurons changes accordingly ($\mathbf{u}_2^\pn$, grey blue, top). As a result, a new signal is observed in the upstream apical dendrite ($\mathbf{v}_{\apic,1}^\pn$, grey bottom) which originates from a mismatch between the top-down feedback (grey blue) and the cancellation given by the lateral interneurons (red). (ii) After $\sim$3s of learning with this new associative signal, plasticity at the bottom-up synapses ($\mathbf{W}_{1,0}^{\pn\pn}$), which receive sensory input ($r_0^\pn$, green), leads to a near-exact prediction of the new, previously unexplained associative signal $\mathbf{u}_2^\mathrm{assoc}$ by $\mathbf{u}_2^\pn$. Consequently, the distal dendrite no longer shows a voltage deflection (error), which results from the top-down and lateral cancellation inputs having the magnitude but opposite signs (blue and red traces, respectively). The network now fully reproduces the new associative signal (top).
  (\textbf{C}) Learning gradually explains away the backpropagated activity. (i) Interneurons learn to predict, and effectively cancel, the backpropagated activity as the lateral weights from the pyramidal-to-interneurons ($\mathbf{W}^{\intn\pn}_{1,1}$) adapt. (ii) While simultaneously adapting the interneuron-to-pyramidal synapses ($\mathbf{W}^{\pn\intn}_{1,1}$), the apical compartment is eventually silenced, even though the pyramidal neurons remain active (not shown). Vertical blue dashed line represents the moment when the new associative signal is presented for the first time.}
  \label{fig:learning_newinput}
\end{figure}

Having established a self-predicting network, we next show how prediction errors get propagated backwards when a novel input is provided to the output area. This new signal, which we model via the activation of additional somatic conductances in output pyramidal neurons (see Methods), plays the role of a teaching or associative signal (see specific tasks below). Here we consider a concrete implementation of the network model introduced above, with an input, a hidden and an output brain area (areas 0, 1 and 2, respectively; Fig.~\ref{fig:learning_newinput}A). We demonstrate learning in the model with a simple task: memorizing a single input-output pattern association. This setup naturally generalizes to multiple memories by iterating over a set of associations to be learned.

When the pyramidal cell activity in the output area is nudged towards some desired target (Fig.~\ref{fig:learning_newinput}B (i)), the bottom-up synapses $\mathbf{W}_{2,1}^{\pn\pn}$ from the lower area neurons to the basal dendrites are adapted, again according to the plasticity rule that implements the dendritic prediction of somatic spiking (see Eq.~\ref{eq:dW-PP} in the Methods and \citetext{Urbanczik2014}). What these synapses cannot explain away shows up as a dendritic error in the pyramidal neurons of the lower area 1. In fact, the self-predicting microcircuit can only cancel the feedback that is produced by the lower area activity. Due to the unexplained teaching signal in the output area, the top-down input partially survives the lateral inhibition; this leads to the activation of distal dendrites (Fig.~\ref{fig:learning_newinput}B (i)). The mathematical analysis reveals that the apical deviation from baseline encodes an error that is effectively backpropagated from the output area.

The somatic integration of apical activity induces plasticity at the bottom-up synapses $\mathbf{W}_{1,0}^{\pn\pn}$ on the basal dendrites. As described above, plasticity at these synapses too is governed by the dendritic prediction of somatic activity, just as for the synapses to the interneurons (Eq.~\ref{eq:dW-PP}). As the apical error changes the somatic activity, plasticity of the $\mathbf{W}_{1,0}^{\pn\pn}$ weights tries to further reduce the error in the output area. Importantly, the plasticity rule depends only on information that is available at the synaptic level. More specifically, it is a function of both postsynaptic firing and dendritic branch voltage, as well as presynaptic activity, in par with detailed phenomenological models \cite{Clopath2010,Bono2017}. In a spiking neuron model, the plasticity rule can reproduce a number of experimental results on spike-timing dependent plasticity \cite{Spicher2017}.

In contrast to the establishing of the self-predicting network state, learning now involves the simultaneous modifications of both lateral circuit and bottom-up synaptic weights (Fig.~\ref{fig:learning_newinput}). On the one hand, lateral weights track changes in output area activity, in this way approximately maintaining the network in a self-predicting state throughout learning. On the other hand, the inputs to area $1$ pyramidal neurons adapt to reduce prediction errors. Altogether, plasticity eventually leads to a network configuration in which the novel top-down input is successfully predicted (Fig.~\ref{fig:learning_newinput}B,C).

\subsection*{Cross-area network learns to solve a nonlinear associative task}

\begin{figure}[htb!]
  \centering
  \includegraphics[width=1\linewidth]{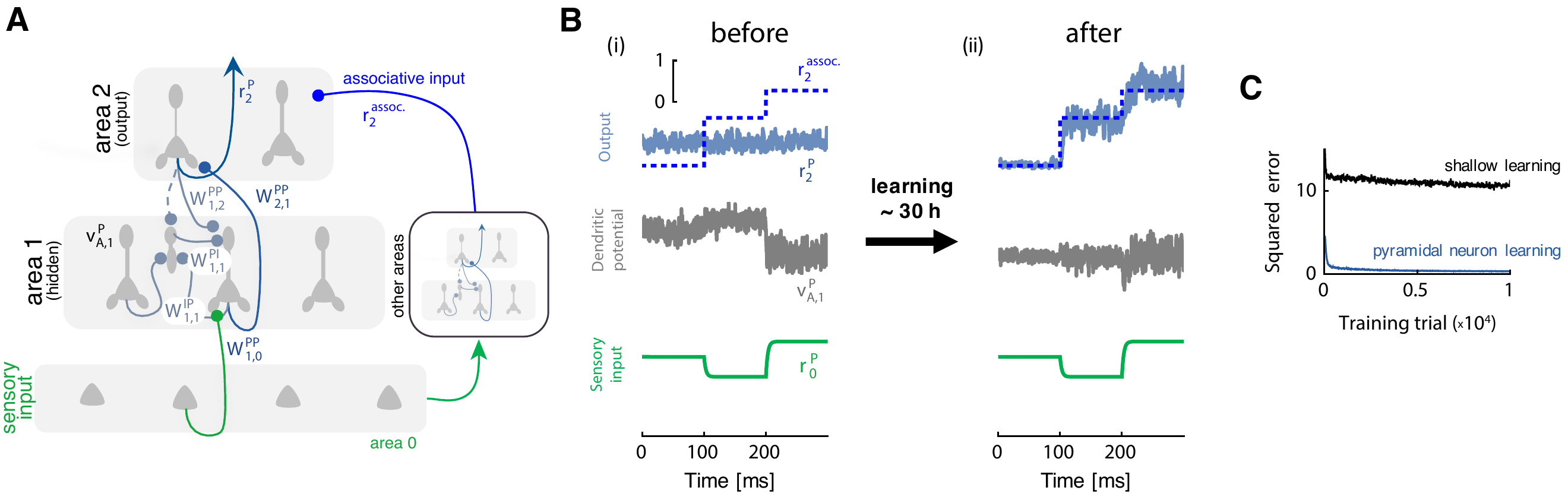}
  \caption{\textbf{A multi-area network learns to solve a nonlinear associative task online in continuous time and without phases.}
  (\textbf{A}) Starting from a self-predicting network state (cf.~Fig.~\ref{fig:learn-self-predicting-state}), a 30-20-10 fully-connected pyramidal neuron network learns to approximate a nonlinear function (represented by the 'other areas' box, another 30-20-10 network) from a stream of input-output pattern pairs. The neuronal and synaptic weight dynamics evolve continuously and without interruption.
  (\textbf{B}) Example firing rates for a randomly chosen output neuron ($r_2^\pn$, blue noisy trace) and its desired target imposed by the associative input ($r_2^\mathrm{assoc}$, blue dashed line),  together with the voltage in the apical compartment of a hidden neuron ($v_{\apic,1}^\pn$, grey noisy trace) and the input rate from the sensory neuron ($r_0^\pn$, green). Before learning (i), the apical dendrite of the hidden neuron shows errors in response to three consecutive input patterns that disappear after 30 h of successful learning (ii). The presentation of the novel output target produces deviations from baseline in the apical compartment, visible in the initial $v_{\apic,1}^\pn$ trace. Such mismatch signals trigger plasticity in the forward weights $\mathbf{W}^{\pn\pn}_{1,0}$, which eventually leads to a reduction of the error in the output area, and henceforth a return to baseline of the apical voltage in the hidden area below.  (\textbf{C}) Error curves for the full model and a shallow learner for comparison, where no backpropagation of errors occurs and only the output weights $\mathbf{W}^{\pn\pn}_{2,1}$ are adapted. }
  \label{fig:learn-function-approximation}
\end{figure}

So far we have described the key components of our model in a multi-area network using a toy problem. Now, we turn to more challenging problems. The first of which is a nonlinear associative task, where the network has to learn to associate the sensory input with the output of a separate multi-area network that transforms the same sensory input --- this can be recast as a nonlinear regression problem (Fig.~\ref{fig:learn-function-approximation}A; see Methods for details on the architecture and learning conditions).

We let learning occur in continuous time without pauses or alternations in plasticity as input patterns are sequentially presented. This is in contrast to previous learning models that rely on computing activity differences over distinct phases, requiring temporally nonlocal computation, or globally coordinated plasticity rule switches  \cite{Hinton1988,OReilly1996,Xie2003,Scellier2017,Guerguiev2017}. Furthermore, we relaxed the bottom-up vs.~top-down weight symmetry imposed by the backprop algorithm and kept the top-down weights $\mathbf{W}_{1,2}^{\pn\pn}$ fixed. Feedback $\mathbf{W}_{1,2}^{\pn\pn}$ weights quickly aligned to $\sim\!45º$ of the forward weights $\left(\mathbf{W}_{2,1}^{\pn\pn}\right)^T$, in line with the recently discovered feedback alignment phenomenon \cite{Lillicrap2016}. This simplifies the architecture, because top-down and interneuron-to-pyramidal synapses need not to be changed. Finally, to test the robustness of the network, we injected a weak noise current to every neuron, as a simple model for uncorrelated background activity (see Methods). Our network was still able to learn this harder task  (Fig.~\ref{fig:learn-function-approximation}B), performing considerably better than a shallow learner where only output weights were adjusted (Fig.~\ref{fig:learn-function-approximation}C). Useful changes were thus made to hidden area $1$ bottom-up weights; the network effectively solved the credit assignment problem.

\subsection*{Multi-area network learns to discriminate handwritten digits}
Next, we turn to a standard machine learning problem, the classification of handwritten digits from the MNIST database. This data set is popularly used to study the performance of learning models, including various artificial neural networks trained with backprop. Notably, shallow models (e.g., logistic regression) or networks trained with plain Hebbian learning alone suffer from poor performance and do not offer a remedy for the problem.

\begin{figure}[htb!]
  \centering
  \includegraphics[scale=1]{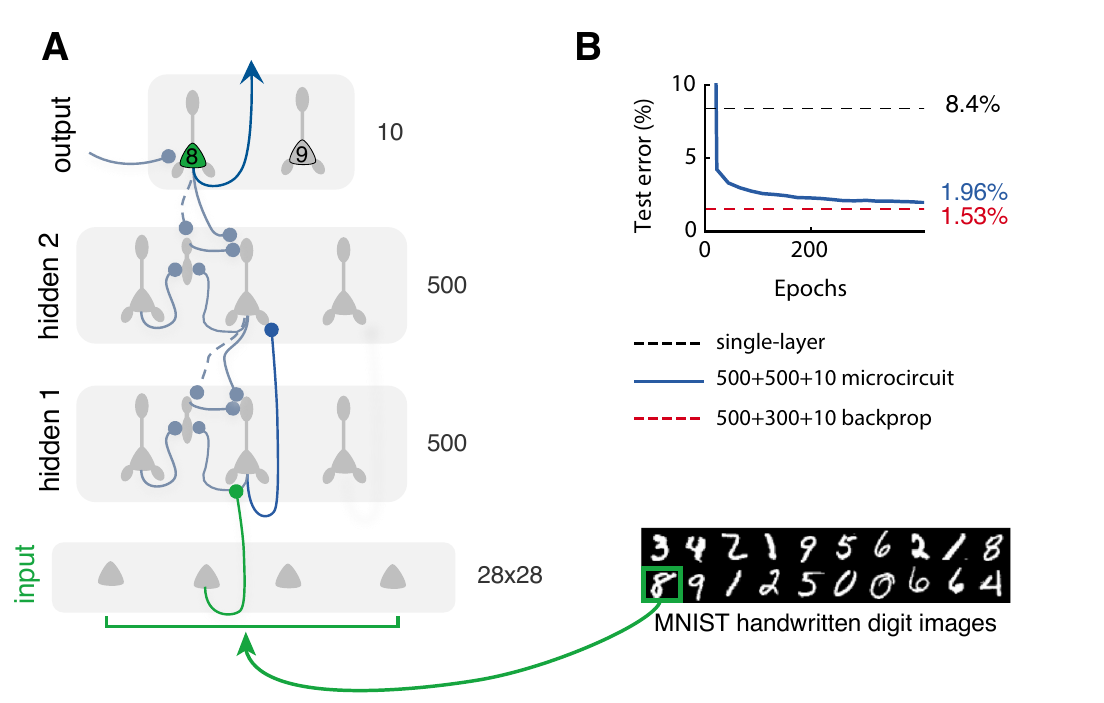}
  \caption{\textbf{Learning to classify real-world, structured stimuli with a multi-area network.}
  (\textbf{A}) A 784-500-500-10 (i.e. with two hidden areas) network of pyramidal neurons learns to recognize and classify handwritten digits from the MNIST data set. Only a subset of connections is shown to enhance clarity.
  (\textbf{B}) Competitive accuracy ($< 2$\%, an empirical signature of backprop-like learning) is achieved on the standard MNIST testing dataset by our network (solid blue). For comparison the performance of a shallow learner (i.e. a network in which only output weights are adapted, dashed black) and of a standard artificial neural network trained with backprop (dashed red, see Methods) are also shown.}
  \label{fig:learn-mnist-classification}
\end{figure}

We wondered how our model would fare in this real-world benchmark, in particular whether the prediction errors computed by the interneuron microcircuit would allow learning the weights of a hierarchical nonlinear network with multiple hidden areas. To that end, we trained a deeper, larger 4-area network (with 784-500-500-10 pyramidal neurons,  Fig.~\ref{fig:learn-mnist-classification}A) by pairing digit images with teaching inputs that nudged the 10 output neurons towards the correct class pattern. To speed up the experiments we studied a simplified network dynamics which determined compartmental potentials without requiring a full neuronal relaxation procedure (see Methods). As in the previous experiments, synaptic weights were randomly initialized and set to a self-predicting configuration where interneurons cancelled top-down inputs, rendering the apical compartments silent before training started. Top-down and interneuron-to-pyramidal weights were kept fixed.

The network was able to achieve a test error of 1.96\%, Fig.~\ref{fig:learn-mnist-classification}B, a figure not overly far from the reference mark of non-convolutional artificial neural networks optimized with backprop (1.53\%) and comparable to recently published results that lie within the range 1.6-2.4\% \cite{Lee2015,Lillicrap2016}. This was possible even though interneurons had to keep track of changes to forward weights as they evolved, simultaneously and without phases. Indeed, apical compartment voltages remained approximately silent when output nudging was turned off (data not shown), reflecting the maintenance of a self-predicting state throughout learning. Moreover, thanks to a feedback alignment dynamics \cite{Lillicrap2016}, the interneuron microcircuit was able to translate the feedback from downstream areas into single neuron prediction error signals, despite the asymmetry of forward and top-down weights and at odds with exact backprop.

\subsection*{Disinhibition enables sensory input generation and sharpening}

So far we assumed that feedback from downstream neurons is relayed through fixed top-down synapses. However, this need not be so. As we demonstrate next, the interneuron microcircuit is capable of tracking changes to the top-down stream dynamically as learning progresses. This endows the model with important additional flexibility, as feedback connections --- known to mediate attention and perceptual acuity enhancement in sensory cortices --- are likely plastic \cite{Huber:2012fw,Petreanu:2012dy,Manita2015,Makino:2015jr,Attinger:2017bo,Leinweber:2017jq}.

\begin{figure}[htb!]
  \centering
  \includegraphics[scale=1]{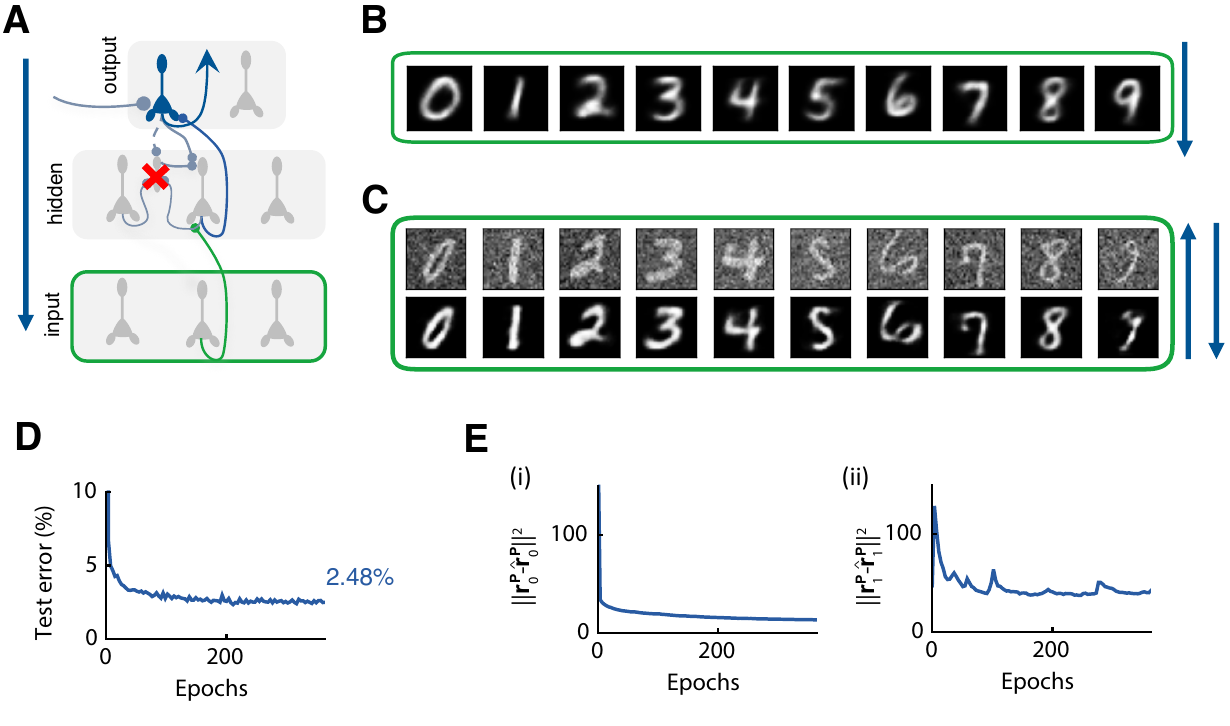}
  \caption{\textbf{Top-down synapses can be adapted to simultaneously drive bottom-up learning, input construction and denoising.}
  (\textbf{A}) Classification performance of a 784-1000-10 network exposed to MNIST images, with plastic top-down synapses that learns to predict lower-area activities. Top-down and forward weights co-adapt without pauses or phases.
  (\textbf{B}) Driving the network top-to-bottom (i.e., initializing the output area to a particular digit and turning off lateral and bottom-up inputs of both hidden and input areas) recreates class-specific image examples in the input area. The top-down connections can be tuned to encode a simple inverse visual model.
  (\textbf{C}) Such an inverse model yields image denoising, which can be achieved by reconstructing corrupted inputs from hidden area activities.
  (\textbf{D})
  The network also successfully learns to classify images.
  (\textbf{E}) Inverse reconstruction losses of original images (i) and hidden (ii) neuron activities. Top-down synapses connecting hidden pyramidal neurons back to the input area learn to reconstruct pixel arrangements given hidden neuron activities; synapses originating at the output area learn to predict hidden area activities given the current class label estimate.
  }
  \label{fig:learn-digit-reconstruction}
\end{figure}

As a case in point we considered a simple extension to a three-area network of 784-1000-10 pyramidal neurons again exposed to MNIST images, Fig.~\ref{fig:learn-digit-reconstruction}. The architecture is as before, except that we now let dendritic predictive plasticity  shape the top-down weights from output to hidden neurons $\mathbf{W}^{\pn\pn}_{1,2}$ as well as an extra set of weights $\mathbf{W}^{\pn\pn}_{0,1}$ connecting hidden neurons back to the input area (see Eq.~\ref{eq:dW-PP-TD} in the Methods).

In this extended network, top-down synapses learn to predict the activities of the corresponding area below and thus implement an approximate inverse of the forward model. In effect, these connections play a dual role, beyond their sole purpose in backprop. They communicate back upper area activities to improve the hidden neuron representation on a recognition task, and they learn to invert the induced forward model. This paired encoder-decoder architecture is known as target propagation in machine learning \cite{Bengio2014,Lee2015}. Our compartmental pyramidal neuron model affords a simple design for the inverse learning paradigm: once more, plasticity of top-down synapses is driven by a postsynaptic dendritic error factor, comparing somatic firing with a local branch potential carrying the current top-down estimate.

Importantly, our results show that the network still learned to recognize handwritten digits, Fig.~\ref{fig:learn-digit-reconstruction}A, reaching a classification error of 2.48\% on the test set. This again highlights that not only transposed forward weight matrices, as prescribed by backprop, deliver useful error signals to hidden areas. In this experiment, we initialized every weight matrix randomly and independently, and did not pre-learn lateral circuit weights. Although forward, top-down and lateral weights were all jointly adapted starting from random initial conditions, a self-predicting state quickly ensued, leading to a drop in classification error. Concomitantly, the reconstructions of hidden neuron activities and input images improved, Fig.~\ref{fig:learn-digit-reconstruction}B.

The learned inverse model can be used to generate prototypical digit images in the input area. We examined qualitatively its performance by directly inspecting the produced images. Specifically, for each digit class we performed a top-to-bottom pass with lateral inhibition turned off, starting from the corresponding class pattern $\mathbf{r}^\pn_2$. For simplicity, we disabled basal feedforward inputs as well to avoid recurrent effects (see Methods). This procedure yielded prototype reconstructions $\hat{\mathbf{r}}^\pn_0 = \phi(\mathbf{W}^{\pn\pn}_{0,1} \, \phi(\mathbf{W}^{\pn\pn}_{1,2} \, \mathbf{r}^\pn_2))$ which resemble natural handwritten digits, Fig.~\ref{fig:learn-digit-reconstruction}C, confirming the observed decrease in reconstruction loss.

Crucially, for the network to be able to generate images, the apical dendrites of hidden neurons should be fully driven by their top-down inputs. In terms of our microcircuit implementation, this is achieved by momentarily disabling the contributions from lateral interneurons. A switch-like disinhibition \cite{Pi:2013ij} is thus capable of turning apical dendrites from error signalling devices into regular prediction units: the generative mode corresponds to a disinhibited mode of operation. Due to their preferential targetting of SST interneurons, VIP interneurons are likely candidates to implement this switch.

Recent reports support the view that cortico-cortical feedback to distal dendrites plays an active role as mice engage in perceptual discrimination tasks \cite{Manita2015,Makino:2015jr,Takahashi2016}. Inspired by these findings, we further tested the capabilities of the model in a visual denoising task, where the prior knowledge incorporated in the top-down network weights is leveraged to improve perception. In Fig.~\ref{fig:learn-digit-reconstruction}D, we show the reconstructions $\hat{\mathbf{r}}_0^\pn = \phi(\mathbf{W}^{\pn\pn}_{0,1} \, \mathbf{r}^\pn_1)$ obtained after presenting randomly picked MNIST images from the test set that had been corrupted with additive Gaussian noise. We show only the apical predictions carried by top-down inputs back to sensory area 0, without actually changing area 0 activity. Interestingly, we found that the hidden neuron representations shaped by classification errors served as reasonable visual features for the inverse model as well. Most of the noise was filtered out, although some of the finer details of the original images were lost in the process.

\section*{Discussion}

How the brain successfully assigns credit and modifies specific synaptic connections given a global associative signal has puzzled neuroscientists for decades. Here we have introduced a novel framework in which a single neuron is capable of transmitting predictions as well as prediction errors. These neuron-specific errors are encoded at distal dendrites and are kept in check by lateral (e.g.~somatostatin-expressing) interneurons. Next, local synaptic plasticity mechanisms use such dendritic-encoded prediction errors to correctly adjust synapses. We have shown that these simple principles allow networks of multiple areas to successfully adjust their weights in challenging tasks, and that this form of learning approximates the well known backpropagation of errors algorithm.

\paragraph{Experimental predictions}
Because our model touches on a number of levels: from brain areas to microcircuits, dendritic compartments and synapses, it makes several predictions. Here we highlight some of these predictions and related experimental observations:

(1) {\em Dendritic error representation.} Probably the most fundamental feature of the model is that dendrites, in particular distal dendrites, encode error signals that instruct learning of lateral and downstream connections. This means that during a task that required the association of two brain areas to develop, lateral interneurons would modify their synaptic weights such that the top-down signals are cancelled. Moreover, during learning, or if this association is broken, a dendritic error signal should be observed. While monitoring such dendritic signals during learning is challenging, there is recent experimental evidence that supports this model. Mice were trained in a simple visuomotor task where the visual flow in a virtual environment presented to the animal was coupled to its own movement \cite{Zmarz:2016jo,Attinger:2017bo}. When this coupling was broken (by stopping the visual flow) mismatch signals were observed in pyramidal cells, consistent with the prediction error signals predicted by our model.

(2) {\em Lateral inhibition of apical activity.} Our apical error representation is based on lateral inhibitory feedback to distal dendritic compartments of pyramidal cells. There is evidence for top-down feedback to target distal (layer-1) synapses of both layer-2/3 and layer-5 pyramidal cells \cite{Petreanu:2009kka}, and both cell types have lateral somatostatin interneurons which target the distal dendrites of the respective pyramidal cells \cite{Markram:2004uv}.
The cancellation of the feedback provided by somatostatin interneurons should be near-exact both in its magnitude and delay. In the brain, there can be a substantial delay between the lateral excitatory input and the feedback from other brain areas (in the order of tens to hundreds of milliseconds \cite{Cauller:1991js,Larkum2013}), suggesting that the lateral inhibitory interaction mediated by SST cells should be also delayed and tuned to the feedback. Interestingly, there is strong experimental support for a delayed inhibition mediated by pyramidal-to-SST connections \cite{Silberberg:2007ila,Murayama2009,Berger:2009gh,Berger:2010js}, which could in principle be tuned to match both the delay and magnitude of the feedback signal. Moreover, the spontaneous activity of SST interneurons is relatively high \cite{UrbanCiecko:2016io}, which again is consistent with our model as SST interneurons need to constantly match the top-down input received by neighbouring pyramidal cells. We would predict that these levels of spontaneous firing rates in SST should match the level of feedback received by the pyramidal cells targeted by a particular SST interneuron. In addition, our model predicts the need for a weak top-down input onto SST interneurons. Again, this is in line with recent top-down connectivity studies suggesting that SST can indeed provide such a precise cancellation of the top-down excitatory inputs \cite{Zhang:2014ib,Leinweber:2017jq}.

(3) {\em Hierarchy of prediction errors} A further implication of our multi-area learning model is that a high-level prediction error occurring at some higher cortical area would imply also lower-level prediction errors to co-occur at earlier areas. For instance, a categorization error occurring when a visual object is misclassified, would be signalled backwards through our interneuron circuits to lower areas where the individual visual features of the objects are represented. Recent experimental observations in the macaque face-processing hierarchy support this view \cite{Schwiedrzik:2017gh}. We predict that higher-area activity modulates lower-area activity with the purpose to shape synaptic plasticity at these lower areas.

\medskip

Here we have focused on the role of SST cells as a feedback-specific interneuron. There are many more interneuron types that we do not consider in our framework. One such type are the PV (parvalbumin-positive) cells, which have been postulated to mediate a somatic excitation-inhibition balance \cite{Vogels2011,Froemke:2015kx} and competition \cite{Masquelier2007,Nessler:2013bu}.
These functions could in principle be combined with the framework introduced here, or as we suggest below, PV interneuron may be involved in representing yet another type of prediction errors different from the classification errors considered so far.
VIP (vasoactive intestinal peptide) interneurons that are believed to be engaged in cortical disinhibition \cite{Letzkus2015} are assumed in our framework to switch between the discriminative mode and the local attention mode in which lower area activity is generated out of higher area activity (see Fig.~\ref{fig:learn-digit-reconstruction}).

We have focused on an interpretation of our predictive microcircuits as learning across brain areas, but they may also be interpreted as learning across different groups of pyramidal cells within the same brain area.

\paragraph{Comparison to previous approaches}
It has been suggested that error backpropagation could be approximated by an algorithm that requires alternating between two learning phases, known as contrastive Hebbian learning \cite{Ackley1985}. This link between the two algorithms was first established for an unsupervised learning task \cite{Hinton1988} and later analyzed \cite{Xie2003} and generalized to a broader class of models \cite{OReilly1996,Scellier2017}. The two phases needed for contrastive Hebbian learning are: (i) for each input pattern, the network first has to settle down being solely driven by inputs; then, (ii) the process is repeated while additionally driving outputs towards a desired target state. Learning requires subtracting activity patterns recorded on each phase --- and therefore requires storing activity in time --- or changing plasticity rules across the network on a coordinated, phase-dependent manner, which appears to be biologically implausible.

Two-phase learning recently reappeared in a study which, like ours, uses compartmental neurons \cite{Guerguiev2017}. In this more recent work, the difference between the activity of the apical dendrite in the presence and the absence of the teaching input represents the error that induces plasticity at the forward synapses. This error is used directly for learning the bottom-up synapses without influencing the somatic activity of the pyramidal cell. In contrast, we postulate that the apical dendrite has an explicit error representation at every moment in time by simultaneously integrating top-down excitation and lateral inhibition. As a consequence, we do not need to postulate separate temporal phases, and our network operates continuously in time while plasticity at all synapses is always turned on.

The solution proposed here to avoid two-phase learning relies on a plastic microcircuit that provides functional lateral inhibition. All the involved plasticity rules are error-correcting in spirit and can be understood as learning to match a certain target voltage. For the synapses from the interneurons to the apical dendrite of the pyramidal neurons, the postsynaptic target is the resting potential, and hence the (functionally) inhibitory plasticity rule can be seen as achieving a dendritic balance, similarly to the homeostatic balance of inhibitory synaptic plasticity as previously suggested (\citenoparens{Vogels2011,Luz2012}). Yet, in our model, inhibitory plasticity plays a central role in multi-area, deep error coding, which goes beyond the standard view of inhibitory plasticity as a homeostatic stabilizing force \cite{Keck2017}.

Error minimization is an integral part of brain function according to predictive coding theories \cite{Rao1999,Friston2005}, and backprop can be mapped onto a predictive coding network architecture \cite{Whittington:2017js}. From a formal point of view this approach is encompassed by the framework introduced by \citetext{LeCun1988}. A possible network implementation is suggested by \citetext{Whittington:2017js} that requires intricate circuitry with appropriately tuned error-representing neurons. According to that model, the only plastic synapses are those that connect prediction and error neurons.

We built upon the previously made observation that top-down and bottom-up weights need not be in perfect agreement to enable multi-area error-driven learning \cite{Lee2015,Lillicrap2016}. Consistent with these findings, the strict weight symmetry arising in the classical error backpropagation algorithm is not required in our case either for a successful learning in hidden area neurons.

\medskip
We have also shown that top-down synapses can be learned using the same dendritic predictive learning rule used at the remaining connections. In our model, the top-down connections have a dual role: they are involved in the apical error representation and, they learn to match the somatic firing driven by the bottom-up input \cite{Urbanczik2014}. The simultaneous learning of the bottom-up and top-down pathways leads to the formation of a generative network that can denoise sensory input or generate dream-like inputs  (Fig.~\ref{fig:learn-digit-reconstruction}).

Finally, the framework introduced here could also be adapted to other types of error-based learning, such as in generative models that instead of learning to discriminate sensory inputs, learn to generate following sensory input statistics. Error propagation in these forms of generative models, which arise from an inaccurate prediction of sensory inputs, may rely on different dendritic compartments and interneurons, such as the previously mentioned PV inhibitory cells \cite{Petreanu:2009kka}.

\section*{Acknowledgements}
The authors would like to thank Timothy P.~Lillicrap, Blake Richards, Benjamin Scellier and Mihai A.~Petrovici for helpful discussions. WS thanks Matthew Larkum for many discussions on dendritic processing and the option of dendritic error representation. In addition, JS thanks Elena Kreutzer, Pascal Leimer and Martin T.~Wiechert for valuable feedback and critical reading of the manuscript.

This work has been supported by the Swiss National Science Foundation (grant 310030L-156863, WS) and the Human Brain Project.

\newpage
\section*{Methods}

\textbf{Neuron and network model.} The somatic membrane potentials of pyramidal neurons and interneurons evolve in time according to
\begin{align}
  \label{eq:dUPdt} \frac{d}{d t}{\mathbf{u}}_k^\pn(t) &= - g_\leak \, \mathbf{u}_k^\pn(t) + g_\bas \left(\mathbf{v}_{\bas, k}^\pn(t) - \mathbf{u}_k^\pn(t) \right) + g_\apic \left(\mathbf{v}_{\apic, k}^\pn(t) - \mathbf{u}_k^\pn(t)\right) + \sigma \, \boldsymbol{\xi}(t)\\
  \label{eq:dUIdt} \frac{d}{d t}{\mathbf{u}}_k^\intn(t)  &= - g_\leak \, \mathbf{u}_k^\intn(t) + g_\dnd \left(\mathbf{v}_{k}^\intn(t) - \mathbf{u}_k^\intn(t) \right) + \mathbf{i}^\intn_k(t) + \sigma \, \boldsymbol{\xi}(t),
\end{align}
with one such pair of dynamical equations for every hidden area $0 < k < N$; input area neurons are indexed by $k = 0$.

Eqs.~\ref{eq:dUPdt} and \ref{eq:dUIdt} describe standard conductance-based voltage integration dynamics, having set membrane capacitance to unity and resting potential to zero for clarity purposes. Background activity is modelled as a Gaussian white noise input, $\boldsymbol{\xi}$ in the equations above. To keep the exposition brief we use matrix notation, and denote by $\mathbf{u}_k^\pn$ and $\mathbf{u}_k^\intn$ the vectors of pyramidal and interneuron somatic voltages, respectively. Both matrices and vectors, assumed column vectors by default, are typed in boldface here and throughout.

As described in the main text, pyramidal hidden neurons are taken as three-compartment neurons to explicitly incorporate basal and apical dendritic integration zones into the model, inspired by the design of L2/3 pyramidal cells. The two dendritic compartments are coupled to the soma with effective transfer conductances $g_\bas$ and $g_\apic$, respectively. Compartmental potentials are given in instantaneous form by
\begin{align}
  \label{eq:Vbas} \mathbf{v}_{\bas, k}^\pn(t) &= \mathbf{W}^{\pn \pn}_{k,k-1} \, \phi(\mathbf{u}_{k-1}^\pn(t)) \\
  \label{eq:Vapic} \mathbf{v}_{\apic, k}^\pn(t) &= \mathbf{W}^{\pn \pn}_{k,k+1} \, \phi(\mathbf{u}_{k+1}^\pn(t)) + \mathbf{W}^{\pn \intn}_{k,k} \, \phi(\mathbf{u}_{k}^\intn(t)),
\end{align}
where $\phi(\mathbf{u})$ is the neuronal transfer function, which acts componentwise on $\mathbf{u}$.

Although the design can be extended to more complex morphologies, in the framework of dendritic predictive plasticity two compartments suffice to compare desired target with actual prediction. Hence, aiming for simplicity, we reduce pyramidal output neurons to two-compartment cells, essentially following \citetext{Urbanczik2014}; the apical compartment is absent ($g_\apic = 0$ in Eq.~\ref{eq:dUPdt}) and basal voltages are as defined in Eq.~\ref{eq:Vbas}. Synapses proximal to the somata of output neurons provide direct external teaching input, incorporated as an additional source of current $\mathbf{i}^\pn_N$. For any given such neuron, excitatory and inhibitory conductance-based input generates a somatic current $i_N^\pn(t) = g_{\exc,N}^\pn(t) \left(E_\exc - u_N^\pn(t)\right) + g_{\inh,N}^\pn(t) \left(E_\inh - u_N^\pn(t)\right)$, where $E_\exc$ and $E_\inh$ are excitatory and inhibitory synaptic reversal potentials, respectively. The point at which no current flows, $i^\pn_N = 0$, defines the target teaching voltage $u_N^\textsf{trgt}$ towards which the neuron is nudged.

Interneurons are similarly modelled as two-compartment cells, cf.~Eq.~\ref{eq:dUIdt}. Lateral dendritic projections from neighboring pyramidal neurons provide the main source of input
\begin{equation}
  \label{eq:Vbasintn} \mathbf{v}_{k}^\intn(t) = \mathbf{W}^{\intn \pn}_{k,k} \, \phi(\mathbf{u}_{k}^\pn(t)),
\end{equation}
whereas cross-area, top-down synapses define the teaching current $\mathbf{i}^\intn_k$. Specifically, an interneuron at area $k$ receives private somatic teaching excitatory and inhibitory input from a pyramidal neuron at area $k\!+\!1$ balanced according to $g_{\exc, k}^\intn = g_\som \frac{u_{k+1}^\pn - E_\inh}{E_\exc - E_\inh}$, $g_{\inh, k}^\intn = - g_\som \frac{u_{k+1}^\pn - E_\exc}{E_\exc - E_\inh}$, where $g_\som$ is some constant scale factor denoting overall nudging strength; with this setting, the interneuron is nudged to follow the corresponding next area pyramidal neuron.

\vspace{1em}
\textbf{Synaptic plasticity.}
Our model synaptic weight update functions belong to the class of dendritic predictive plasticity rules \cite{Urbanczik2014,Spicher2017} that can be expressed in general form as
\begin{equation}
  \label{eq:dwdt}
  \frac{d}{dt} w = \eta \, h(v) \left(\phi(u) - \phi(v) \right) r,
\end{equation}
where $w$ is an individual synaptic weight, $\eta$ is a learning rate, $u$ and $v$ denote distinct compartmental potentials, $\phi$ is a rate function, third factor $h$ is a function of potential $v$, and $r$ is the presynaptic input. Eq.~\ref{eq:dwdt} was originally derived in the light of reducing the prediction error of somatic spiking, when $u$ represents the somatic potential and $v$ is a function of the postsynaptic dendritic potential.

Concretely, the plasticity rules for the various connection types present in the network are:
\begin{align}
  \label{eq:dW-PP}\frac{d}{dt} \mathbf{W}^{\pn\pn}_{k,k-1} &= \eta^{\pn\pn}_{k,k-1} \left(\phi(\mathbf{u}^\pn_k) - \phi(\hat{\mathbf{v}}^{\pn}_{\bas,k}) \right) \left(\mathbf{r}_{k-1}^\pn\right)^{\! T},\\
  \label{eq:dW-IP}\frac{d}{dt} \mathbf{W}^{\intn\pn}_{k,k} &= \eta^{\intn\pn}_{k,k} \left(\phi(\mathbf{u}^\intn_k) - \phi(\hat{\mathbf{v}}^{\intn}_{k}) \right) \left(\mathbf{r}_{k}^\pn\right)^{\! T},\\
  \label{eq:dW-PI}\frac{d}{dt} \mathbf{W}^{\pn\intn}_{k,k} &= \eta^{\pn\intn}_{k,k} \left(\mathbf{v}_\rest - \mathbf{v}^{\pn}_{\apic,k} \right) \left(\mathbf{r}_{k}^\intn\right)^{\! T},
\end{align}
where $(\cdot)^T$ denotes vector transpose and $\mathbf{r}_k \equiv \phi(\mathbf{u}_k)$ the area $k$ firing rates. So the strengths of plastic synapses evolve according to the correlation of dendritic prediction error and presynaptic rate and can undergo both potentiation or depression depending on the sign of the first factor.

For basal synapses, such prediction error factor amounts to a difference between postsynaptic rate and a local dendritic estimate which depends on the branch potential. In Eqs.~\ref{eq:dW-PP} and ~\ref{eq:dW-IP}, dendritic predictions $\hat{\mathbf{v}}^{\pn}_{\bas,k} = \frac{g_\bas}{g_\leak + g_\bas + g_\apic} \, \mathbf{v}^{\pn}_{\bas,k}$ and $\hat{\mathbf{v}}^{\intn}_{k} = \frac{g_\dnd}{g_\leak + g_\dnd} \, \mathbf{v}^{\intn}_{k}$ take into account dendritic attenuation factors. Meanwhile, plasticity rule \eqref{eq:dW-PI} of lateral interneuron-to-pyramidal synapses aims to silence (i.e., set to resting potential $\mathbf{v}_\rest = \mathbf{0}$, here and throughout null for simplicity) the apical compartment; this introduces an attractive state for learning where the contribution from interneurons balances top-down dendritic input. The learning rule of apical-targetting synapses can be thought of as a dendritic variant of the homeostatic inhibitory plasticity proposed by \citetext{Vogels2011}.

In the experiments where the top-down connections are plastic (cf.~Fig.~\ref{fig:learn-digit-reconstruction}), the weights evolve according to
\begin{equation}
  \label{eq:dW-PP-TD}\frac{d}{dt} \mathbf{W}^{\pn\pn}_{k,k+1} = \eta^{\pn\pn}_{k,k+1} \left(\phi(\mathbf{u}^\pn_k) - \phi(\hat{\mathbf{v}}_{\topdown,k}^\pn) \right) \left(\mathbf{r}_{k+1}^\pn\right)^{\! T},
\end{equation}
with $\hat{\mathbf{v}}_{\topdown,k}^\pn = \mathbf{W}_{k,k+1} \, \mathbf{r}_{k+1}^{\pn}$.
An implementation of this rule requires a subdivision of the apical compartment into a distal part receiving the top-down input (with voltage $\hat{\mathbf{v}}_{\topdown,k}^\pn$) and a more proximal part receiving the lateral input from the interneurons (with voltage $\mathbf{v}_{\apic,k}^\pn)$.

\vspace{1em}
\textbf{Nonlinear function approximation task.} In Fig.~\ref{fig:learn-function-approximation}, a pyramidal neuron network learns to approximate a random nonlinear function implemented by a held-aside feedforward network with the same (30-20-10) dimensions; this ensures that the target function is realizable. One teaching example consists of a randomly drawn input pattern $\mathbf{r}_0^\pn$ assigned to corresponding target $\mathbf{r}_2^\mathrm{trgt} = \phi(\mathbf{W}^\mathrm{trgt}_{2,1} \, \phi(\mathbf{W}^\mathrm{trgt}_{1,0} \, \mathbf{r}_0^\pn))$. Teacher network weights and input pattern entries are sampled from a uniform distribution $U(-1,1)$. We choose a soft rectifying nonlinearity as the neuronal transfer function, $\phi(u) = \log(1 + \exp(u))$.

The pyramidal neuron network is initialized to a self-predicting state where $\mathbf{W}^{\intn\pn}_{1,1} = \mathbf{W}^{\pn\pn}_{2,1}$ and $\mathbf{W}^{\pn\intn}_{1,1} = - \mathbf{W}^{\pn\pn}_{1,2}$. Top down weight matrix $\mathbf{W}^{\pn\pn}_{1,2}$ is fixed and set at random with entries drawn from a uniform distribution. Output area teaching currents $\mathbf{i}_2^\pn$ are set so as to nudge $\mathbf{u}_2^\pn$ towards the teacher-generated $\mathbf{u}_2^\mathrm{trgt}$. Reported error curves are exponential moving averages of the sum of squared errors loss $\| r_{2}^\pn - r_{2}^\mathrm{trgt} \|^2$ computed after every example on unseen input patterns. Plasticity induction terms given by the right-hand sides of Eqs.~\ref{eq:dW-PP}-\ref{eq:dW-PI} are low-pass filtered with time constant $\tau_w$ before being definitely consolidated, to dampen fluctuations; synaptic plasticity is kept on throughout. Plasticity and neuron model parameters are given in the accompanying supplementary material.

\vspace{1em}
\textbf{MNIST image classification and reconstruction tasks.}  When simulating the larger models used on the MNIST data set we resort to a discrete-time network dynamics where the compartmental potentials are updated in two steps before applying synaptic changes.

The simplified model dynamics is as follows. For each presented MNIST image, both pyramidal and interneurons are first initialized to their bottom-up prediction state \eqref{eq:Vbas}, $\mathbf{u}_k = \mathbf{v}_{\bas,k}$, starting from area $1$ upto top area $N$. Output area neurons are then nudged towards their desired target $\mathbf{u}_N^\mathrm{trgt}$, yielding updated somatic potentials $\mathbf{u}^\pn_N = (1 - \lambda_N) \, \mathbf{v}_{\bas,N} + \lambda_N \, \mathbf{u}^\mathrm{trgt}_N$. To obtain the remaining final compartmental potentials, the network is revisited in reverse order, proceeding from area $k=N-1$ down to $k=1$. For each $k$, interneurons are first updated to include top-down teaching signals, $\mathbf{u}_k^\intn = (1 - \lambda_I) \, \mathbf{v}_k^\intn + \lambda_I \, \mathbf{u}_{k+1}^\pn$; this yields apical compartment potentials according to \eqref{eq:Vapic}, after which we update hidden area somatic potentials as a convex combination with mixing factor $\lambda_k$. The convex combination factors introduced above are directly related to neuron model parameters as conductance ratios. Synaptic weights are then updated according to Eqs.~\ref{eq:dW-PP}-\ref{eq:dW-PP-TD}.

Such simplified dynamics approximates the full recurrent network relaxation in the deterministic setting $\sigma \to 0$, with the approximation improving as the top-down dendritic coupling is decreased, $g_\apic \to 0$.

We train the models on the standard MNIST handwritten image database, further splitting the training set into 55000 training and 5000 validation examples. The reported test error curves are computed on the 10000 held-aside test images. The four-area network shown in Fig.~\ref{fig:learn-mnist-classification} is initialized in a self-predicting state with appropriately scaled initial weight matrices. To speed-up training we use a mini-batch strategy on every learning rule, whereby weight changes are averaged across 10 images before being actually consolidated. We take the neuronal transfer function $\phi$ to be a logistic function, $\phi(u) = 1/(1 + \exp(-u))$ and include a learnable threshold on each neuron, modelled as an additional input fixed at unity with plastic weight. Desired target class vectors are 1-hot coded, with $r_N^\mathrm{trgt} \in \{0.1, 0.8\}$. During testing, the output is determined by picking the class label corresponding to the neuron with highest firing rate. Model parameters are given in full in the supplementary material.

To generate digit prototypes as shown in Fig.~\ref{fig:learn-digit-reconstruction}C, the network is ran feedforward in a top-to-bottom fashion: a pass of pyramidal neuron activations is performed, while disabling the feedforward stream as well as the interneuron negative lateral contributions. For this reason, this mode of recall is referred to in the main text as the disinhibited mode. The output area is initialized to the 1-hot-coded pattern corresponding to the desired digit class.

The denoised images shown in Fig.~\ref{fig:learn-digit-reconstruction}D are the top-down predictions $\hat{\mathbf{r}}_0 = \phi(\hat{\mathbf{v}}^\pn_{\mathrm{TD},0})$ obtained after presenting randomly selected digit examples from the test set, corrupted with additive Gaussian noise of standard deviation $\sigma=0.3$. The network states are determined by the two-step procedure described above. Recurrent effects are therefore ignored, as a single backward step is performed.

\vspace{1em}
\textbf{Computer code.} For the first series of experiments (Figs.~\ref{fig:learn-self-predicting-state}-\ref{fig:learn-function-approximation}) we wrote custom Mathematica (Wolfram Research, Inc.) code. The larger MNIST networks (Figs.~\ref{fig:learn-mnist-classification} and \ref{fig:learn-digit-reconstruction}) were simulated in Python using the TensorFlow framework.

\bibliographystyle{jneurosci}
\bibliography{pyramidal}

\newpage
\beginsupplement

\section*{Supplementary information}

\subsection*{Supplementary data}
Below we detail the model parameters used to generate the figures presented in the main text.

\vspace{1em}
\textbf{Fig.~\ref{fig:learn-self-predicting-state} details}. The parameters for the compartmental model neuron were: $g_\apic = 0.8$, $g_\bas = g_\dnd = 1.0$, $g_\leak = 0.1$. Interneuron somatic teaching conductances were balanced to yield overall nudging strength $g_\som = 0.8$. Initial weight matrix entries were independently drawn from a uniform distribution $U(-1, 1)$. We chose background activity levels of $\sigma=0.1$. The learning rates were set as $\eta^{\intn\pn}_{1,1} = 0.0002375$ and $\eta^{\pn\intn}_{1,1} = 0.0005$.

Input patterns were smoothly transitioned by low-pass filtering $\mathbf{u}_0^\pn$ with time constant $\tau_0 = 3$. A transition between patterns was triggered every 100 ms. Weight changes were low pass filtered with time constant $\tau_w = 30$. The dynamical equations were solved using Euler's method with a time step of 0.1, which resulted in 1000 integration time steps per pattern.

\vspace{1em}
\textbf{Fig.~\ref{fig:learning_newinput} details}. We used learning rates $\eta_{1,0}^{\pn\pn} = \eta_{1,1}^{\intn\pn} = 0.0011875$ and $\eta_{2,1}^{\pn\pn} = 0.0005$. Remaining parameters as used for Fig.~\ref{fig:learn-self-predicting-state}.

\vspace{1em}
\textbf{Fig.~\ref{fig:learn-function-approximation} details}. Initial forward weights $\mathbf{W}_{1,0}^{\pn\pn}$ and $\mathbf{W}_{2,1}^{\pn\pn}$ were scaled down by a factor of 0.1. Background noise level was raised to $\sigma = 0.3$. The learning rates were $\eta^{\intn\pn}_{1,1} = 0.00002375$, $\eta^{\pn\pn}_{1,0} = 0.00011875$, $\eta^{\pn\pn}_{2,1} = 0.00001$. Weight matrices $\mathbf{W}^{\pn\pn}_{1,2}$ and $\mathbf{W}^{\pn\intn}_{1,1}$ were kept fixed, so the model relied on a feedback alignment mechanism to learn. Remaining parameters as used for Fig.~\ref{fig:learn-self-predicting-state}.

\vspace{1em}
\textbf{Fig.~\ref{fig:learn-mnist-classification} details}. We chose mixing factors $\lambda_3 = \lambda_I = 0.1$ and $\lambda_1 = \lambda_2 = 0.3$. Forward learning rates were $\eta_{3,2}^{\pn\pn} = 0.001/\lambda_3$, $\eta_{2,1}^{\pn\pn} = \eta_{3,2}^{\pn\pn}/\lambda_2$, $\eta_{1,0}^{\pn\pn} = \eta_{2,1}^{\pn\pn}/\lambda_1$. Lateral learning rates were $\eta_{2,2}^{\intn\pn} = 2 \eta_{3,2}^{\pn\pn}$ and $\eta_{1,1}^{\intn\pn} = 2 \eta_{2,1}^{\pn\pn}$. Initial forward weights were drawn at random from a uniform distribution $U(-0.1, 0.1)$, and the remaining weights from $U(-1,1)$.

\vspace{1em}
\textbf{Fig.~\ref{fig:learn-digit-reconstruction} details}. We took all mixing factors equal $\lambda_2 = \lambda_1 = \lambda_I = 0.1$. Forward learning rates: $\eta_{2,1}^{\pn\pn} = 0.02/\lambda_2$, $\eta_{1,0}^{\pn\pn} = \eta_{2,1}^{\pn\pn}/\lambda_1$. Lateral connections learned with rate $\eta_{1,1}^{\intn\pn} = \eta_{1,1}^{\pn\intn} = \eta_{2,1}^{\pn\pn}$. Top-down connections were initialized from a uniform distribution $U(-0.1, 0.1)$ and adapted with learning rates $\eta_{1,2}^{\pn\pn} = 0.0002$ and $\eta_{0,1}^{\pn\pn} = 0.0001$.

\subsection*{Supplementary analysis}
In this supplementary note we present a set of mathematical results concerning the network and plasticity model described in the main text.

To proceed analytically we make a number of simplifying assumptions. Unless noted otherwise, we study the network in a deterministic setting and consider the limiting case where lateral microcircuit synaptic weights match the corresponding forward weights:
\begin{align}
  \label{eq:ideal-wpi} \mathbf{W}^{\pn \intn}_{k, k} &= - \mathbf{W}_{k, k+1}^{\pn \pn} \equiv \mathbf{W}^{\pn \intn*}_{k,k}\\
  \label{eq:ideal-wip}\mathbf{W}^{\intn \pn}_{k, k} &= \frac{g_\bas + g_\leak}{g_\bas + g_\apic + g_\leak} \mathbf{W}_{k+1, k}^{\pn \pn} \equiv \mathbf{W}^{\intn \pn*}_{k,k},
\end{align}
The particular choice of proportionality factors, which depend on the neuron model parameters, is motivated below. Under the above configuration, the network becomes self-predicting.

To formally relate the encoding and propagation of errors implemented by the inhibitory microcircuit to the backpropagation of errors algorithm from machine learning, we consider the limit where top-down input is weak compared to the bottom-up drive. This limiting case results in error signals that decrease exponentially with area depth, but allows us to proceed analytically.

We further assume that the top-down weights converging to the apical compartments are equal to the corresponding forward weights, $\mathbf{W}^{\pn\pn}_{k, k+1} = \left(\mathbf{W}^{\pn\pn}_{k+1, k} \right)^T$. Such weight symmetry is not essential for successful learning in a broad range of problems, as demonstrated in the main simulations and as observed before \cite{Lee2015,Lillicrap2016}. It is, however, required to frame learning as a gradient descent procedure. Furthermore, in the analyses of the learning rules, we assume that synaptic changes take place at a fixed point of the neuronal dynamics; we therefore consider discrete-time versions of the plasticity rules. This approximates the continuous-time plasticity model as long as changes in the inputs are slow compared to the neuronal dynamics.

For convenience, we will occasionally drop neuron type indices and refer to bottom-up weights $\mathbf{W}_{k+1, k}$ and to top-down weights $\mathbf{W}_{k, k+1}$. Additionally, we assume without loss of generality that the dendritic coupling conductance for interneurons is equal to the basal dendritic coupling of pyramidal neurons, $g_\dnd = g_\bas$. Finally, whenever it is useful to distinguish whether output area nudging is turned off, we use superscript `$-$'.

\vspace{1em}
\textbf{Interneuron activity in the self-predicting state.}
Following \citetext{Urbanczik2014}, we note that steady state interneuron somatic potentials can be expressed as a convex combination of basal dendritic and pyramidal neuron potentials that are provided via somatic teaching input:
\begin{equation}
\label{eq:intn-convcomb}
  \mathbf{u}_k^{\intn} =  \frac{g_\bas}{g_\leak + g_\bas + g_\som} \, \mathbf{v}_{k}^{\intn} + \frac{g_\som}{g_\leak + g_\bas + g_\som} \, \mathbf{u}_{k+1}^{\pn} = (1 - \lambda) \,  \hat{\mathbf{v}}_{k}^{\intn} + \lambda \, \mathbf{u}_{k+1}^{\pn},
\end{equation}
with $g_\bas$ and $g_\leak$ the effective dendritic transfer and leak conductances, respectively, and $g_\som$ the total excitatory and inhibitory teaching conductance. In the equation above, $\hat{\mathbf{v}}_{k}^{\intn} = \frac{g_\bas}{g_\leak + g_\bas} \mathbf{v}_{k}^{\intn}$ is the interneuron dendritic prediction (cf.~Eq.~\ref{eq:dW-IP}), and $\lambda \equiv \frac{g_\som}{g_\leak + g_\bas + g_\som} \in [0,1[$ is a mixing factor which controls the nudging strength for the interneurons. In other words, the current prediction $\hat{\mathbf{v}}_{k}^{\intn}$ and the teaching signal are averaged with coefficients determined by normalized conductances. We will later consider the weak nudging limit of $\lambda \to 0$.

The relation $\hat{\mathbf{v}}_{k}^{\intn} = \hat{\mathbf{v}}_{\bas,k+1}^{\pn}$ holds when pyramidal-to-interneuron synaptic weights are equal to pyramidal-pyramidal forward weights, up to a scale factor: $\mathbf{W}^{\intn\pn}_{k,k} = \frac{g_\leak + g_\bas}{g_\leak + g_\bas + g_\apic} \mathbf{W}^{\pn\pn}_{k+1,k}$, which simplifies to $\mathbf{W}^{\intn\pn}_{N-1,N-1} = \mathbf{W}^{\pn\pn}_{N,N-1}$ for the last area where $g_\apic=0$ (to reduce clutter, we use the slightly abusive notation whereby $g_\apic$ should be understood to be zero when referring to output area neurons). This is the reason for the particular choice of ideal pyramidal-to-interneuron weights presented in the preamble. The network is then internally consistent, in the sense that the interneurons predict the model's own predictions, held by pyramidal neurons.

\vspace{1em}
\textbf{Bottom-up predictions in the absence of external nudging.}
We first study the situation where the input pattern $\mathbf{r}_0$ is stationary and the output area teaching input is disabled, $\mathbf{i}^\pn_{N} = 0$. We show that the fixed point of the network dynamics is a state where somatic voltages are equal to basal voltages, up to a dendritic attenuation factor. In other words, the network effectively behaves as if it were feedforward, in the sense that it computes the same function as the corresponding network with equal bottom-up but no top-down or lateral connections.

Specifically, in the absence of external nudging (indicated by the $-$ in the superscript), the somatic voltages of pyramidal and interneuron are given by the bottom-up dendritic predictions,
\begin{align}
  \label{eq:FF-state}
  \mathbf{u}^{\pn,-}_k = \hat{\mathbf{v}}_{\bas,k}^{\pn,-} &\equiv \frac{g_\bas}{g_\leak + g_\bas + g_\apic} \, \mathbf{W}^{\pn \pn}_{k, k-1} \, \phi(\hat{\mathbf{v}}_{\bas, k-1}^{\pn,-})\\
  \mathbf{u}^{\intn,-}_k = \hat{\mathbf{v}}_{k}^{\intn,-} &\equiv \frac{g_\bas}{g_\leak + g_\bas} \, \mathbf{W}^{\intn \pn}_{k, k} \, \phi(\hat{\mathbf{v}}_{\bas, k}^{\pn,-}).
\end{align}

To show that Eq.~\ref{eq:FF-state} describes the state of the network, we start at the output area and set Eq.~\ref{eq:dUPdt} to zero. Because nudging is turned off, we observe that $\mathbf{u}_N^{\pn}$ is equal to $\hat{\mathbf{v}}_{\bas, N}^{\pn, -}$ if area $N-1$ also satisfies $\mathbf{u}_{N-1}^{\pn} = \hat{\mathbf{v}}_{\bas, N-1}^{\pn, -}$. The same recursively applies to the hidden area below when its apical voltage vanishes, $\mathbf{v}_{\apic, N-1}^\pn = 0$. Now we note that at the fixed point the interneuron cancels the corresponding pyramidal neuron, due to the assumption that the network is in a self-predicting state, which yields $\mathbf{u}_{N-1}^{\intn} = \mathbf{u}_{N}^{\pn}$. Together with the fact that $\mathbf{W}^{\pn \intn}_{N-1, N-1} = - \mathbf{W}_{N-1, N}^{\pn \pn}$, we conclude that the interneuron contribution to the apical compartment cancels the top-down pyramidal neuron input, yielding the required condition $\mathbf{v}_{\apic, N-1}^\pn = 0$.

The above argument can be iterated down to the input area, which is constant, and we arrive at Eq.~\ref{eq:FF-state}.

\vspace{1em}
\textbf{Zero plasticity induction in the absence of nudging.} In view of Eq.~\ref{eq:FF-state}, which states that in the absence of external nudging the somatic voltages correspond to the basal predictions, no synaptic changes are induced in basal synapses on the pyramidal and interneurons as defined by the plasticity rules \eqref{eq:dW-PP} and \eqref{eq:dW-IP}, respectively. Similarly, the apical voltages are equal to rest, $\mathbf{v}^{\pn,-}_{\apic,k}=\mathbf{v}_\rest$, when the top-down input is fully predicted, and no synaptic plasticity is induced in the inter-to-pyramidal neuron synapses, see \eqref{eq:dW-PI}. When noisy background currents are present, the average prediction error is zero, while momentary fluctuations will still trigger plasticity. Note that the above holds when the dynamics is away from equilibrium, under the additional constraint that the integration time constant of interneurons matches that of pyramidal neurons.

\vspace{1em}
\textbf{Recursive prediction error propagation.}
Prediction errors arise in the model whenever lateral interneurons cannot fully explain top-down input, leading to a deviation from baseline in apical dendrite activity. Here, we look at the network steady state equations for a stationary input pattern $\mathbf{r}_0$ and derive an iterative relationship which establishes the propagation across the network of prediction mismatches originating downstream. The following compartmental potentials are thus evaluated at a fixed point of the neuronal dynamics.

Under the assumption \eqref{eq:ideal-wpi} of matching interneuron-to-pyramidal top-down weights, apical compartment potentials simplify to
\begin{equation}
\label{eq:v-apic}
  \mathbf{v}^\pn_{\apic,k} = \mathbf{W}_{k, k+1} \left[\phi(\mathbf{u}^\pn_{k+1}) - \phi(\mathbf{u}^\intn_{k})\right] = \mathbf{W}_{k, k+1} \, \mathbf{e}_{k+1},
\end{equation}
where we introduced error vector $\mathbf{e}_{k+1}$ defined as the difference between pyramidal and interneuron firing rates. Such deviation can be intuitively understood as an area-wise interneuron prediction mismatch, being zero when interneurons perfectly explain pyramidal neuron activity. We now evaluate this difference vector at a fixed point to obtain a recurrence relation that links consecutive areas.

The steady-state somatic potentials of hidden pyramidal neurons are given by
\begin{align}
\label{eq:pn-star}
  \mathbf{u}_{k}^{\pn} &=  \frac{g_\bas}{g_\leak + g_\bas + g_\apic} \, \mathbf{v}_{\bas,k}^{\pn} + \frac{g_\apic}{g_\leak + g_\bas + g_\apic} \, \mathbf{v}_{\apic,k}^{\pn} = \hat{\mathbf{v}}_{\bas,k}^{\pn} + \lambda \, \mathbf{v}_{\apic,k}^{\pn}\nonumber\\
  &= \hat{\mathbf{v}}_{\bas,k}^{\pn} + \lambda \, \mathbf{W}_{k, k+1} \, \mathbf{e}_{k+1}.
\end{align}
To shorten the following, we assumed that the apical attenuation factor is equal to the interneuron nudging strength $\lambda$. As previously mentioned, we proceed under the assumption of weak feedback, $\lambda$ small. As for the corresponding interneurons, we insert Eq.~\ref{eq:pn-star} into Eq.~\ref{eq:intn-convcomb} and note that when the network is in a self-predicting state we have $\hat{\mathbf{v}}_{k-1}^{\intn} = \hat{\mathbf{v}}_{\bas,{k}}^{\pn}$, yielding
\begin{equation}
\label{eq:intn-star}
  \mathbf{u}_{k-1}^{\intn} =  (1 - \lambda) \,  \hat{\mathbf{v}}_{\bas,k}^{\pn} + \lambda \left(\hat{\mathbf{v}}_{\bas,k}^{\pn} + \lambda \, \mathbf{v}_{\apic,k}^{\pn} \right) = \hat{\mathbf{v}}_{\bas,k}^{\pn} +  \lambda^2 \, \mathbf{v}_{\apic,k}^{\pn}.
\end{equation}

Using the identities \eqref{eq:pn-star} and \eqref{eq:intn-star}, we now expand to first order the difference vector $\mathbf{e}_{k}$ around $\hat{\mathbf{v}}_{\bas,k}^{\pn}$ as follows
\begin{align}
  \mathbf{e}_{k} &= \phi(\mathbf{u}_{k}^\pn) - \phi(\mathbf{u}_{k-1}^\intn) = \lambda  \, \mathbf{D}_{k} \, \mathbf{v}_{\apic,k}^{\pn} + \mathcal{O}\!\left(\lambda^2 \, \| \mathbf{v}_{\apic,k}^\pn \|\right).
\end{align}
Matrix $\mathbf{D}_{k}$ is a diagonal matrix with diagonal equal to $\phi^\prime(\hat{\mathbf{v}}_{\bas,k}^{\pn})$, i.e., whose $i$-th element reads $\frac{d\phi}{dv} (\hat{v}_{\bas,k,i}^{\pn})$. It contains the derivative of the neuronal transfer function $\phi$ evaluated component-wise at the bottom-up predictions $\hat{\mathbf{v}}_{\bas,k+1}^{\pn}$. Recalling Eq.~\ref{eq:v-apic}, we obtain a recurrence relation
\begin{align}
  \mathbf{e}_{k} &= \lambda \, \mathbf{D}_{k} \, \mathbf{W}_{k, k+1} \, \mathbf{e}_{k+1} + \mathcal{O}\!\left(\lambda^2 \, \| \mathbf{W}_{k,k+1} \, \mathbf{e}_{k+1} \|\right).
\end{align}

Finally, last area pyramidal neurons provide the initial condition by being directly nudged towards the desired target  $\mathbf{u}_N^\textsf{trgt}$. Their membrane potentials can be written as
\begin{equation}
  \mathbf{u}_{N}^{\pn} = (1 - \lambda) \, \hat{\mathbf{v}}_{\bas,N}^{\pn} + \lambda \, \mathbf{u}_N^\textsf{trgt},
\end{equation}
and this gives an estimate for the error in the output area of the form
\begin{equation}
  \label{eq:delta-K}
  \mathbf{e}_{N} = \lambda \, \mathbf{D}_{N} \left(\mathbf{u}_N^\textsf{trgt} - \hat{\mathbf{v}}_{\bas,N}^{\pn}\right) + \mathcal{O}\!\left(\lambda^2 \, \|\mathbf{u}_N^\textsf{trgt} - \hat{\mathbf{v}}_{\bas,N}^{\pn} \|\right) ,
\end{equation}
where for simplicity we took the same mixing factor $\lambda$ for pyramidal output and interneurons. Then, for an arbitrary area, assuming that the synaptic weights and the remaining fixed parameters do not scale with $\lambda$, we arrive at
\begin{equation}
  \label{eq:delta-star}
  \mathbf{e}_{k} = \lambda^{N-k+1} \, \left(\prod_{l=k}^{N-1} \mathbf{D}_{l} \, \mathbf{W}_{l, l+1} \right) \, \mathbf{D}_{N} \left(\mathbf{u}_N^\textsf{trgt} - \hat{\mathbf{v}}_{\bas,N}^{\pn}\right) + \mathcal{O}(\lambda^{N-k+2}).
\end{equation}

Thus, steady state potentials of apical dendrites (cf.~Eq.~\ref{eq:v-apic}) recursively encode neuron-specific prediction errors that can be traced back to a mismatch at the output area.

\vspace{1em}
\textbf{Learning as approximate error backpropagation.}
In the previous section we found that neurons implicitly carry and transmit error information across the network. We now show how the proposed synaptic plasticity model, when applied at a steady state of the neuronal dynamics, can be recast as an approximate gradient descent learning procedure.

More specifically, we compare our model against learning through backprop \cite{Rumelhart1986} or approximations thereof \cite{Lee2015,Lillicrap2016} the weights of the feedfoward multi-area network obtained by removing interneurons and top-down connections from the intact network. For this reference model, the activations $\mathbf{u}_k^-$ are by construction equal to the bottom-up predictions obtained in the full model when output nudging is turned off, $\mathbf{u}_k^- \equiv \hat{\mathbf{v}}_{\bas,k}^{\pn,-}$, cf.~Eq.~\ref{eq:FF-state}. Thus, optimizing the weights in the feedforward model is equivalent to optimizing the predictions of the full model.

Define the loss function
\begin{equation}
  \mathcal{L} \! \left(\mathbf{u}_{N}^-, \mathbf{u}_{N}^\textsf{trgt} \right) = - \sum_{i=1}^{N_N} \int_0^{u_{N,i}^-} \phi\left((1-\lambda) \, \nu + \lambda \, u^\textsf{trgt}_{N,i}\right) - \phi(\nu) \, d\nu,
\end{equation}
where $N_N$ denotes the number of output neurons. $\mathcal{L}$ can be thought of as the multi-area, multi-output unit analogue of the loss function optimized by the single neuron model \cite{Urbanczik2014}, where it stems directly from the particular chosen form of the learning rule \eqref{eq:dW-PP}. The nudging strength parameter $\lambda \in [0, 1[$ allows controlling the mixing with the target and can be understood as an additional learning rate parameter. Albeit unusual in form, function $\mathcal{L}$ imposes a cost similar to an ordinary squared error loss. Importantly, it has a minimum when $\mathbf{u}^{-}_N = \mathbf{u}_N^\textsf{trgt}$ and it is lower bounded. Furthermore, it is differentiable with respect to compartmental voltages (and synaptic weights). It is therefore suitable for gradient descent optimization. As a side remark, $\mathcal{L}$ integrates to a quadratic function when $\phi$ is linear.

Gradient descent proceeds by changing synaptic weights according to
\begin{equation}
\Delta \mathbf{W}_{k,k-1} = - \eta \, \frac{\partial \mathcal{L}}{\partial \mathbf{W}_{k,k-1}}.
\end{equation}
The required partial derivatives can be efficiently computed by the backpropagation of errors algorithm. For the network architecture we study, this yields a learning rule of the form
\begin{equation}
  \label{eq:Delta-W-bp}
   \Delta \mathbf{W}_{k,k-1}^\textsf{bp} = \eta \, \mathbf{e}_{k}^- \, \phi(\mathbf{u}^-_{k-1})^T.
\end{equation}
The error factor $\mathbf{e}_{k}^-$ can be expressed recursively as follows:
\begin{equation}
  \mathbf{e}_{k}^- =
  \begin{cases}
    \phi \! \left((1-\lambda) \, \mathbf{u}_N^- + \lambda \, \mathbf{u}_N^\textsf{trgt} \right) - \phi \!\left(\mathbf{u}_N^-\right) & \text{ if } k = N,\\
    \mathbf{D}^-_k \, \mathbf{W}^T_{k+1,k} \, \mathbf{e}^-_{k+1} & \text{ otherwise,}
  \end{cases}
\end{equation}
ignoring constant factors that depend on conductance ratios, which can be dealt with by redefining learning rates or backward pass weights. As in the previous section, matrix $\mathbf{D}_k^-$ is a diagonal matrix, with diagonal equal to $\phi^\prime(\mathbf{u}^-_k)$.

We first compare the fixed point equations of the original network to the feedforward activations of the reference model. Starting from the bottommost hidden area, using Eqs.~\ref{eq:v-apic}, \ref{eq:pn-star} and \ref{eq:delta-star}, we notice that $\mathbf{u}_1^{\pn} = \mathbf{u}_1^{-} + \lambda \, \mathbf{v}_{\apic,1}^{\pn} = \mathbf{u}_1^{-} + \mathcal{O}(\lambda^N)$, as the bottom-up input is the same in both cases. Inserting this into second hidden area steady state potentials and linearizing the neuronal transfer function gives $\mathbf{u}_2^{\pn} = \mathbf{u}_2^{-} + \lambda \, \mathbf{v}_{\apic,2}^{\pn} + \mathcal{O}(\lambda^{N}) =  \mathbf{u}_2^{-} + \mathcal{O}(\lambda^{N-1})$. This can be repeated and for an arbitrary area and neuron type we find
\begin{align}
  \label{eq:pn-minus-star} \mathbf{u}_k^{\pn} &= \mathbf{u}_k^-  + \lambda \, \mathbf{v}_{\apic,k}^{\pn} + \mathcal{O}(\lambda^{N-k+2}) = \mathbf{u}_k^- + \mathcal{O}(\lambda^{N-k+1})\\
  \label{eq:intn-minus-star}\mathbf{u}_{k-1}^{\intn} &= \mathbf{u}_k^- + \mathcal{O}(\lambda^{N-k+2}).
\end{align}
Writing Eq.~\ref{eq:pn-minus-star} in the first form emphasizes that the apical contributions dominate $\mathcal{O}(\lambda \, \mathbf{v}_{\apic,k}^{\pn}) = \mathcal{O}(\lambda^{N-k+1})$ the bottom-up corrections, which are of order $\mathcal{O}(\lambda^{N-k+2})$.

Next, we prove that up to a factor and to first order the apical term in Eq.~\ref{eq:pn-minus-star} represents the backpropagated error in the feedforward network, $\mathbf{e}_{k}^-$. Starting from the topmost hidden area apical potentials, we reevaluate difference vector \eqref{eq:delta-K} using \eqref{eq:pn-minus-star}. Linearization of the neuronal transfer function gives
\begin{equation}
  \mathbf{v}_{\apic,N-1}^{\pn} =  \lambda \, \mathbf{W}_{N-1,N} \, \mathbf{D}^-_N \left(\mathbf{u}_N^\textsf{trgt} - \mathbf{u}_N^- \right) + \mathcal{O}(\lambda^2).
\end{equation}
Inserting the expression above into Eq.~\ref{eq:pn-minus-star} and using Eq.~\ref{eq:intn-minus-star} the apical compartment potentials at area $N-1$ can then be recomputed. This procedure can be iterated until the input area is reached. In general form, somatic membrane potentials at hidden area $k$ can be expressed as
\begin{align}
  \mathbf{u}^{\pn}_k &= \mathbf{u}_k^- + \lambda \, \mathbf{v}_{\apic,k}^{\pn} + \mathcal{O}(\lambda^{N-k+2})\\
  &= \mathbf{u}_k^- + \lambda^{N-k+1} \, \mathbf{W}_{k,k+1} \, \left(\prod_{l=k+1}^{N-1} \mathbf{D}_{l}^- \, \mathbf{W}_{l, l+1} \right) \mathbf{D}_{N}^- \left(\mathbf{u}_N^\textsf{trgt} - \mathbf{u}_N^- \right) + \mathcal{O}(\lambda^{N-k+2}).
\end{align}
This equation shows that, to leading order of $\lambda$, hidden neurons mix and propagate forward purely bottom-up predictions with top-down errors that are computed at the output area and spread backwards.

We are now in position to compare model synaptic weight updates to the ones prescribed by backprop. Output area updates are exactly equal by construction, $\Delta \mathbf{W}_{N,N-1} = \Delta \mathbf{W}^\textsf{bp}_{N,N-1}$. For pyramidal-to-pyramidal neuron synapses from hidden area $k-1$ to area $k$, we obtain
\begin{align}
  \Delta \mathbf{W}_{k,k-1} &= \eta_{k,k-1} \left[ \phi(\mathbf{u}_{k}^{\pn}) - \phi(\hat{\mathbf{v}}_{\bas,k}^{\pn}) \right] \left(\mathbf{r}_{k-1}^{\pn}\right)^T \nonumber\\
  &= \eta_{k,k-1} \left[\phi\left(\mathbf{u}_{k}^- +  \lambda \, \mathbf{v}_{\apic,k}^{\pn} + \mathcal{O}(\lambda^{N-k+2})\right) - \phi(\mathbf{u}_{k}^-)\right] \left(\mathbf{r}_{k-1}^- + \mathcal{O}(\lambda^{N-k+2}) \right)^T \nonumber\\
  \label{eq:Delta-W-weak} &= \eta_{k,k-1} \lambda^{N-k+1} \left(\prod_{l=k}^{N-1} \mathbf{D}_{l}^- \, \mathbf{W}_{l, l+1} \right) \mathbf{D}_{N}^- \left(\mathbf{u}_N^\textsf{trgt} - \mathbf{u}_N^- \right) \left(\mathbf{r}_{k-1}^-\right)^T + \mathcal{O}(\lambda^{N-k+2}),
\end{align}
while backprop learning rule \eqref{eq:Delta-W-bp} can be written as
\begin{align}
  \Delta \mathbf{W}_{k,k-1}^\textsf{bp} &= \eta \, \lambda \, \left(\prod_{l=k}^{N-1} \mathbf{D}_{l}^- \, \mathbf{W}_{l, l+1} \right) \mathbf{D}_{N}^- \left(\mathbf{u}_N^\textsf{trgt} - \mathbf{u}_N^- \right) \left(\mathbf{r}_{k-1}^-\right)^T + \mathcal{O}(\lambda^2),
\end{align}
where we used that, to first order, the output area error factor is $\mathbf{e}_N^- = \lambda \, \mathbf{D}^-_N \left(\mathbf{u}_N^\textsf{trgt} - \mathbf{u}_N^- \right) + \mathcal{O}(\lambda^2)$. Hence, up to a factor of $\lambda^{N-k}$ which can be absorbed in the learning rate $\eta_{k,k-1}$, changes induced by synaptic plasticity are equal to the backprop learning rule \eqref{eq:Delta-W-bp} in the limit $\lambda \to 0$, provided that the top-down weights are set to the transpose of the corresponding feedforward weights, $\mathbf{W}_{k,k+1} = \mathbf{W}_{k+1,k}^T$. The `quasi-feedforward' condition $\lambda \to 0$ has also been invoked to relate backprop to two-phase contrastive Hebbian learning in Hopfield networks \cite{Xie2003}.

In our simulations, top-down weights are either set at random and kept fixed, in which case Eq.~\ref{eq:Delta-W-weak} shows that the plasticity model optimizes the predictions according to an approximation of backprop known as feedback alignment \cite{Lillicrap2016}; or learned so as to minimize an inverse reconstruction loss, in which case the network implements a form of difference target propagation \cite{Lee2015}.

\vspace{1em}
\textbf{Interneuron plasticity.}
The analyses of the previous sections relied on the assumption that the synaptic weights to and from interneurons were set to their ideal values, cf.~Eqs.~\ref{eq:ideal-wpi} and \ref{eq:ideal-wip}. We now study the plasticity of the lateral microcircuit synapses and show that, under mild conditions, learning rules \eqref{eq:dW-IP} and \eqref{eq:dW-PI} yield the desired synaptic weight matrices.

We first study the learning of pyramidal-to-interneuron synapses $\mathbf{W}^{\intn\pn}_{k,k}$. To quantify the degree to which the weights deviate from their optimal setting, we introduce the convex loss function
\begin{equation}
  \mathcal{L}_k^{\intn\pn} = \frac{1}{2} \Tr \left\{ (\mathbf{W}^{\intn\pn*}_{k,k} - \mathbf{W}^{\intn\pn}_{k,k} )^T (\mathbf{W}^{\intn\pn*}_{k,k} - \mathbf{W}^{\intn\pn}_{k,k} ) \right\},
\end{equation}
where $\Tr(\mathbf{M})$ denotes the trace of matrix $\mathbf{M}$ and $\mathbf{W}^{\intn\pn*}_{k,k} = \frac{g_\bas + g_\leak}{g_\bas + g_\apic + g_\leak} \mathbf{W}_{k+1, k}^{\pn \pn}$, as defined in Eq.~\ref{eq:ideal-wip}.

Starting from the pyramidal-to-interneuron synaptic plasticity rule \eqref{eq:dW-IP}, we express the interneuron somatic potential in convex combination form \eqref{eq:intn-convcomb} and then expand to first order around $\hat{\mathbf{v}}_k^\intn$,
\begin{align}
  \Delta \mathbf{W}_{k,k}^{\intn\pn} &= \eta^{\intn\pn}_{k,k} \left(\phi(\mathbf{u}_k^\intn) - \phi(\hat{\mathbf{v}}_k^\intn) \right) \, (\mathbf{r}^\pn_k)^T \nonumber\\
  &= \eta^{\intn\pn}_{k,k} \, \lambda \, \mathbf{D}_k^{\intn\pn} \left(\mathbf{u}_{k+1}^\pn - \hat{\mathbf{v}}_k^\intn \right) (\mathbf{r}^\pn_k)^T + \mathcal{O}(\lambda^2) \nonumber\\
  &= \eta^{\intn\pn}_{k,k} \, \lambda \, \frac{g_\bas}{g_\leak + g_\bas} \, \mathbf{D}_k^{\intn\pn} \left(\mathbf{W}_{k,k}^{\intn\pn*} - \mathbf{W}_{k,k}^{\intn\pn} \right) \mathbf{Q}_k + \mathcal{O}(\lambda^2) + \mathcal{O}(\lambda \, \alpha).
\end{align}
Matrix $\mathbf{Q}_k = \mathbf{r}_k^\pn \, (\mathbf{r}_k^\pn)^T$ denotes the outer product, and $\mathbf{D}_k^{\intn\pn}$ is a diagonal matrix with $i$-th diagonal entry equal to $\phi^\prime(\hat{\mathbf{v}}_{k,i}^{\intn})$.

For simplicity, we ignore fluctuations arising from the stochastic sequential presentation of patterns \cite{Bottou1998} and look only at the expected synaptic dynamics\footnote{This can be understood as a batch learning protocol, where weight changes are accumulated in the limit of many patterns before being effectively consolidated as a synaptic update.}. We absorb irrelevant scale factors and to avoid a vanishing update we rescale the learning rate $\hat{\eta}^{\intn\pn}_{k,k}$ by $\lambda^{-1}$. Then, taking the limit $\lambda \to 0$, $\alpha \to 0$ as in the previous sections yields
\begin{align}
  \Expect \! \left[\Delta \mathbf{W}_{k,k}^{\intn\pn} \right] &= \hat{\eta}^{\intn\pn}_{k,k} \left(\mathbf{W}_{k,k}^{\intn\pn*} - \mathbf{W}_{k,k}^{\intn\pn} \right) \Expect \! \left[ \mathbf{D}_k^{\intn\pn} \mathbf{Q}_k \right]\nonumber\\
  &= - \hat{\eta}^{\intn\pn}_{k,k} \, \frac{\partial \mathcal{L}_k^{\intn\pn}}{\partial \mathbf{W}^{\intn\pn}_{k,k}} \, \Expect \! \left[ \mathbf{D}_k^{\intn\pn} \mathbf{Q}_k \right] \label{eq:expected-dW-IP}.
\end{align}
The expectation is taken over the pattern ensemble. In the last equality above, we used the fact that the gradient of $\mathcal{L}_k^{\intn\pn}$ with respect to the lateral weights $\mathbf{W}^{\intn\pn}_{k,k}$ is given by the difference $\mathbf{W}^{\intn\pn*}_{k,k} - \mathbf{W}^{\intn\pn}_{k,k}$.

As long as the expectation on the right-hand side of \eqref{eq:expected-dW-IP} is positive definite, the synaptic dynamics is within $90º$ of the gradient and thus leads to the unique minimum of $\mathcal{L}_k^{\intn\pn}$. This condition is easily met in practice. For linear neurons, it amounts to requiring that the correlation matrix $\Expect [\mathbf{Q}]$ is positive definite. In other words, the patterns have to span $\mathbb{R}^{N_k}$, with $N_k$ being the number of pyramidal neurons at area $k$. This is fulfilled when uncorrelated background noise currents are present, and it is likely the case for deterministic networks solving nontrivial tasks. For nonlinear neurons with a monotonically increasing transfer function $\phi$, saturation can lead to a matrix numerically close to singular and slow down learning. A weight matrix initialization that sets the neurons operating far from saturation is therefore an appropriate choice.

A mathematical analysis of the coupled system defined by the various plasticity rules acting in concert is rather involved and beyond the scope of this note. However, the learning of apical-targetting interneuron-to-pyramidal synapses can be studied in isolation by invoking a separation of timescales argument. To proceed, we assume that pyramidal-to-interneuron synapses are ideally set, $\mathbf{W}_{k,k}^{\intn\pn} = \mathbf{W}_{k,k}^{\intn\pn*}$. In practice, this translates to a choice of a small effective learning rate for apical-targetting weights $\eta_{k,k}^{\pn\intn}$. Note that, indirectly, this requirement also imposes a constraint on how fast top-down pyramidal-to-pyramidal weights $\mathbf{W}_{k,k+1}^{\pn\pn}$ can evolve. This is the parameter regime explored in the main text simulations with plastic top-down weights, Fig.~\ref{fig:learn-digit-reconstruction}.

We can then proceed in a similar manner to the previous analysis, as we briefly outline below. Recalling from Eq.~\ref{eq:ideal-wpi} that $\mathbf{W}^{\pn\intn*}_{k,k} = - \mathbf{W}^{\pn\pn}_{k,k+1}$, we define the loss function
\begin{equation}
  \mathcal{L}_k^{\pn\intn} = \frac{1}{2} \Tr \left\{ (\mathbf{W}^{\pn\intn*}_{k,k} - \mathbf{W}^{\pn\intn}_{k,k} )^T (\mathbf{W}^{\pn\intn*}_{k,k} - \mathbf{W}^{\pn\intn}_{k,k} ) \right\}.
\end{equation}
After some manipulation, as $\lambda \to 0$ the expected synaptic change can be written as
\begin{equation}
  \label{eq:expected-dW-PI} \Expect \! \left[\Delta \mathbf{W}_{k,k}^{\pn\intn} \right] = - \eta_{k,k}^{\pn\intn} \, \frac{\partial \mathcal{L}_k^{\pn\intn}}{\partial{\mathbf{W}_{k,k}^{\pn\intn}}} \,
  \Expect \! \left[ \mathbf{r}^\pn_{k+1} \, (\mathbf{r}_{k+1}^\pn)^T \right],
\end{equation}
which leads us to conclude that the weights converge to the appropriate values, provided that the correlation matrix of area $k\!+\!1$ activity patterns is positive definite.

\end{document}